\documentclass[lettersize,journal]{IEEEtran}
\usepackage{amsmath,amsfonts}
\usepackage{algorithmic}
\usepackage{algorithm}
\usepackage{array}
\usepackage[caption=false,font=normalsize,labelfont=sf,textfont=sf]{subfig}
\usepackage{textcomp}
\usepackage{stfloats}
\usepackage{url}
\usepackage{verbatim}
\usepackage{graphicx}
\usepackage{cite}
\usepackage[utf8]{inputenc}
\usepackage[T1]{fontenc}
\usepackage{hyperref}
\usepackage{url}
\usepackage{booktabs}
\usepackage{amsfonts}
\usepackage{nicefrac}
\usepackage{microtype}
\usepackage{xcolor}
\usepackage{bm}
\usepackage{amsmath}
\usepackage{comment}
\usepackage{float}
\usepackage{tabularx}
\usepackage{bbm}
\usepackage{multirow}
\usepackage{lipsum}
\usepackage{pifont}
\usepackage{amssymb}
\usepackage{color, colortbl}
\usepackage{tikz}
\usepackage{enumitem}
\usepackage{xfrac}

\newcommand{\cmark}{\ding{51}}
\newcommand{\xmark}{\ding{55}}

\def\C{{\mathbb{C}}}

\def\diag{{\mathsf{diag}}}

\def\xbmhat{{\widehat{\bm{x}}}}

\def\xbmtilde{{\widetilde{\bm{x}}}}

\def\zbmhat{{\widehat{\bm{z}}}}

\def\kbm{{\bm{k}}}

\def\mbm{{\bm{m}}}
\def\nbm{{\bm{n}}}

\def\pbm{{\bm{p}}}
\def\qbm{{\bm{q}}}

\def\xbm{{\bm{x}}}
\def\ybm{{\bm{y}}}
\def\zbm{{\bm{z}}}

\def\Ebm{{\bm{E}}}
\def\Fbm{{\bm{F}}}

\def\Mbm{{\bm{M}}}

\def\Rbm{{\bm{R}}}
\def\Sbm{{\bm{S}}}

\def\thetabm{{\bm{\theta}}}
\def\phibm{{\bm{\phi}}}

\def\tackle{{\textsc{Tackle}}}
\def\loupefov{{LOUPE$_\text{FOV}$}}
\def\louperoi{{LOUPE$_\text{ROI}$}}
\def\louperecon{{LOUPE$_\text{recon.}$}}
\def\tacklefov{{\textsc{Tackle}$_\text{FOV}$}}
\def\tackleroi{{\textsc{Tackle}$_\text{ROI}$}}
\def\tacklerecon{{\textsc{Tackle}$_\text{recon.}$}}
\def\tackleseg{{\textsc{Tackle}$_\text{seg.}$}}
\def\tackleclass{{\textsc{Tackle}$_\text{class.}$}}

\newcommand{\lblapp}[1]{\label{app:#1}}
\newcommand{\lblsec}[1]{\label{sec:#1}}
\newcommand{\lblfig}[1]{\label{fig:#1}} 
\newcommand{\lbltab}[1]{\label{tbl:#1}}

\newcommand{\lbleq}[1]{\label{eq:#1}}
\newcommand{\refsec}[1]{Section~\ref{sec:#1}}
\newcommand{\reffig}[1]{Figure~\ref{fig:#1}} 
\newcommand{\reftab}[1]{Table~\ref{tbl:#1}}

\newcommand{\refeq}[1]{\eqref{eq:#1}}

\newcommand{\resetcounter}{
  \setcounter{figure}{10} 
  \setcounter{table}{6} 
}

\begin{document}

\title{Learning Task-Specific Strategies \\ for Accelerated MRI}

\author{Zihui Wu,~\IEEEmembership{Graduate Student~Member,~IEEE}, Tianwei Yin, Yu Sun,~\IEEEmembership{Member,~IEEE}, Robert Frost \\ Andre van der Kouwe,~\IEEEmembership{Senior~Member,~IEEE}, Adrian V. Dalca, Katherine L. Bouman
\thanks{This work was sponsored by NSF Award 2048237, NIH Projects 5R01AG064027, 5R01AG070988, R21EB029641, R01HD099846, R01HD085813, Heritage Medical Research Fellowship, S2I Clinard Innovation Award, and Rockley Photonics. Z. Wu was sponsored by the Kortschak Fellowship, Amazon AI4Science Fellowship, and Amazon AI4Science Partnership Discovery Grant.}
\thanks{Z. Wu, Y. Sun, and K. L. Bouman are with the Department of Computing and Mathematical Sciences, California Institute of Technology, Pasadena, CA 91105, USA (email: zwu2@caltech.edu; sunyu@caltech.edu; klbouman@caltech.edu).}
\thanks{T. Yin and A. V. Dalca are with the Computer Science and Artificial Intelligence Lab (CSAIL), Massachusetts Institute of Technology, Cambridge, MA 02139, USA (email: tianweiy@mit.edu; adalca@mit.edu).}
\thanks{R. Frost, A. van der Kouwe, and A. V. Dalca are with the Athinoula A. Martinos Center for Biomedical Imaging, Department of Radiology, MGH, Harvard Medical School, Charlestown, MA 02129, USA (email: srfrost@mgh.harvard.edu; avanderkouwe@mgh.harvard.edu; adalca@mit.edu).}
\thanks{This paper has supplementary downloadable material available at \href{http://ieeexplore.ieee.org}{http://ieeexplore.ieee.org}, provided by the authors. The material includes additional implementation details and experimental results. This material is 1.5 MB in size.}
\thanks{More information is available at \href{http://imaging.cms.caltech.edu/tackle/}{http://imaging.cms.caltech.edu/tackle/}.}
}

\markboth{IEEE Transactions on Computational Imaging, VOL. 10, 2024}
{Wu \MakeLowercase{\textit{et al.}}: Learning Task-Specific Strategies for Accelerated MRI}

\maketitle

\begin{abstract}
Compressed sensing magnetic resonance imaging (CS-MRI) seeks to recover visual information from subsampled measurements for diagnostic tasks.
Traditional CS-MRI methods often separately address measurement subsampling, image reconstruction, and task prediction, resulting in a suboptimal end-to-end performance.
In this work, we propose \tackle~as a unified co-design framework for jointly optimizing subsampling, reconstruction, and prediction strategies for the performance on downstream tasks.
The na\"ive approach of simply appending a task prediction module and training with a task-specific loss leads to suboptimal downstream performance.
Instead, we develop a training procedure where a backbone architecture is first trained for a generic pre-training task (image reconstruction in our case), and then fine-tuned for different downstream tasks with a prediction head.
Experimental results on multiple public MRI datasets show that \tackle~achieves an improved performance on various tasks over traditional CS-MRI methods.
We also demonstrate that \tackle~is robust to distribution shifts by showing that it generalizes to a new dataset we experimentally collected using different acquisition setups from the training data.
Without additional fine-tuning, \tackle~leads to both numerical and visual improvements compared to existing baselines.
We have further implemented a learned 4$\times$-accelerated sequence on a Siemens 3T MRI Skyra scanner. 
Compared to the fully-sampling scan that takes 335 seconds, our optimized sequence only takes 84 seconds, achieving a four-fold time reduction as desired, while maintaining high performance.
Our code is available at \href{https://github.com/zihuiwu/TACKLE}{https://github.com/zihuiwu/TACKLE}.
\end{abstract}


\begin{figure}[t]
\centering
\includegraphics[width=0.45\textwidth]{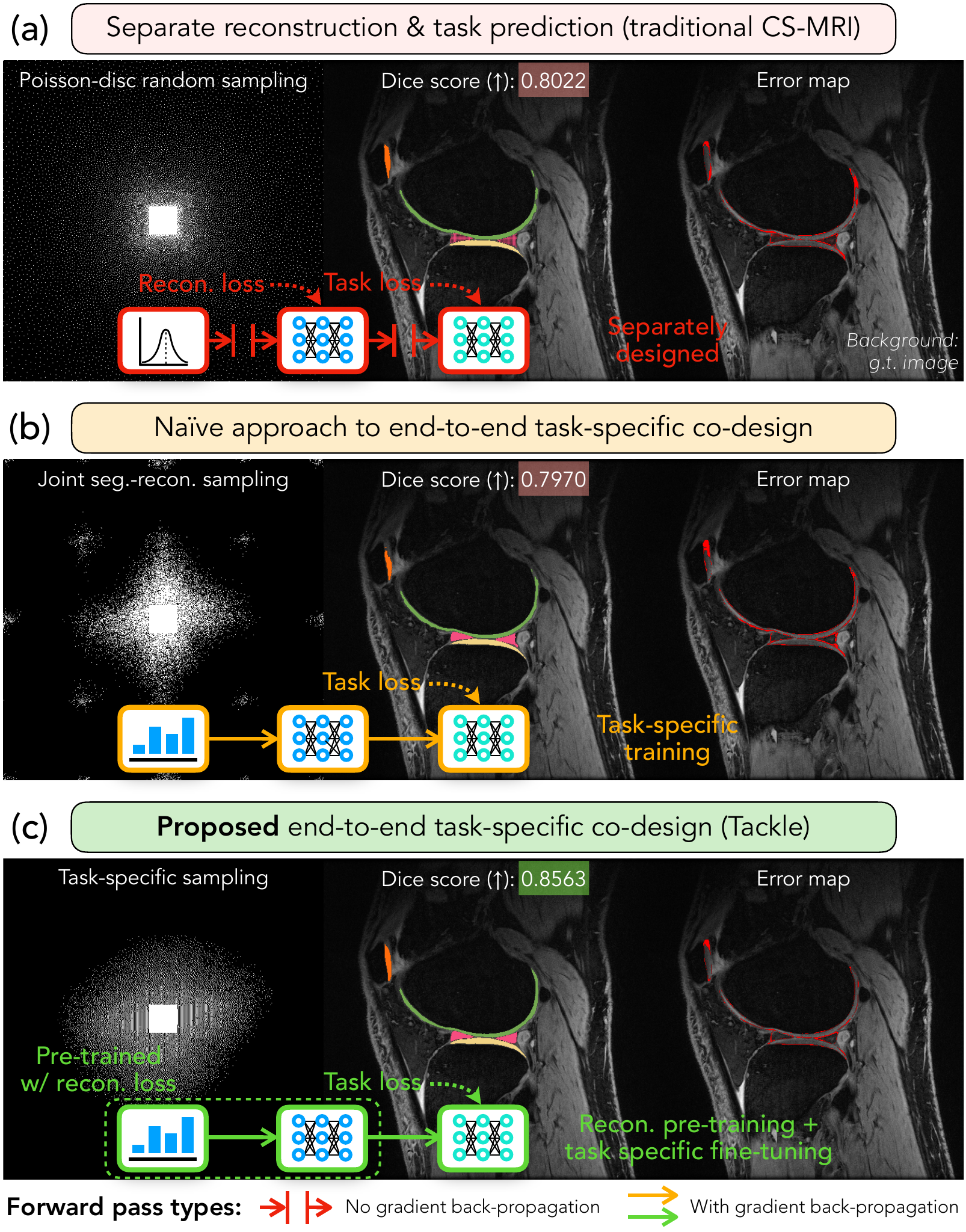}
\caption{
Comparison between (a) traditional CS-MRI, (b) a na\"ive approach to task-specific CS-MRI, and (c) the proposed \tackle~framework.
Compared with panel (a) that separately deals with reconstruction and task prediction, panel (b) is a simple extension of co-design methods for solving downstream tasks by adding a learnable mapping from measurements to task predictions.
However, this na\"ive approach leads to a suboptimal performance and can even lead to a worse task prediction accuracy, as shown in the example above.
On the other hand, we introduce \tackle~for effectively learning task-specific CS-MRI strategies.
\tackle~is first pre-trained for generic reconstruction, and then all three modules are fine-tuned for a more specific downstream task.
We find that this training schedule allows \tackle~to robustly learn generalizable task-specific strategies.
In the above knee segmentation example, all three approaches are trained with the same architectures for the reconstructor (second module) and predictor (third module).
Nevertheless, \tackle~significantly outperforms the two baseline approaches.
}
\vspace{-10pt}
\lblfig{teaser}
\end{figure}

\section{Introduction}
Compressed sensing magnetic resonance imaging (CS-MRI) is a popular accelerated MRI technology~\cite{lustig2007sparse}.
Commonly, CS-MRI is formulated as an imaging inverse problem where the goal is to recover a high-quality image from its subsampled measurements. 
Traditional CS-MRI techniques include solving a regularized optimization problem~\cite{lustig2008compressed, romano2016red, ahmad2020plugandplay, liu2020rare} or training deep learning (DL) models~\cite{yang2016deep, schlemper2018deep, zhang2018istanet} that recover an image from a pre-determined set of measurements.
Recently, a new group of DL-based methods, known as \emph{co-design}, has been proposed to jointly optimize the choice of measurements and a reconstruction module, leading to better reconstruction performance than the traditional CS-MRI methods~\cite{bahadir2019learning, zhang2019reducing, zhang2020extending, aggarwal2020jmodl, alkan2020learning, wang2021bspline, xue20222d, peng2022learning, wang2022stochastic, yang2022fast, martinini2022deep, wang2022joint, wang2022leaders, chaithya2022hybrid, zibetti2022alternating}.

In the existing co-design literature, task prediction is often viewed as a post-processing step decoupled from image reconstruction.
All the aforementioned methods focus on image reconstruction and rely on standard image similarity metrics such as mean square error (MSE) or peak signal-to-noise ratio (PSNR) as a proxy for performance on a downstream task. 
Such a reconstruction-oriented formulation lacks a direct connection with the downstream tasks that reflect actual clinical needs~\cite{menze2014multimodal}.
We are thus motivated to ask: \emph{can one improve the accuracy of downstream task prediction by optimizing the entire CS-MRI pipeline in an end-to-end fashion?}

With end-to-end co-design methods, it seems like we are only one step away from incorporating downstream tasks as part of the optimization. 
Namely, one can simply append a task prediction module and add a task-specific loss.
However, as shown in \reffig{teaser} and \reftab{exp_skmtea}, this approach leads to a suboptimal performance on the task prediction and is sometimes even worse than the traditional approach of separate reconstruction and task prediction.
These results indicate that it remains a challenge on how to robustly learn task-specific strategies for CS-MRI.

In this paper, we address this challenge by proposing a unified framework, \emph{\textbf{ta}sk-specific \textbf{c}odesign of $\bm{k}$-space subsamp\textbf{l}ing and pr\textbf{e}diction} (\tackle), for designing task-specific CS-MRI systems. 
Different from existing works that focus on specific tasks, \tackle~is a general framework that accommodates different downstream tasks.
To do so, we design a two-step training strategy that mimics the training of modern language and vision models.
\tackle~is first trained for a generic task of image reconstruction, and then fine-tuned for specific downstream tasks.
We find that this approach can effectively learn generalizable task-specific strategies that lead to significant and consistent improvements, with an example shown in \reffig{teaser} (c). 
Besides the standard task of reconstructing the full field-of-view (which we call full-FOV reconstruction hereafter), we demonstrate \tackle~on three other tasks covering both pixel-level and image-level imaging problems: region-of-interest (ROI) oriented reconstruction, tissue segmentation, and pathology classification.
Our experimental results show that end-to-end optimization for task prediction sometimes circumvents the typical reconstruction in terms of pixel-wise accuracy, but leads to improved accuracy on the task of interest by effectively extracting key visual information for task prediction.

The main contributions of this work are as follows:
\begin{itemize}
    \item We provide a general framework (\tackle) that learns specific strategies for a variety of CS-MRI tasks. \tackle~optimizes the entire CS-MRI pipeline, from measurement acquisition to label prediction, in an end-to-end fashion \emph{directly} for a user-defined task.
    \item We validate \tackle~on multiple MRI datasets, covering different body parts, scanning sequences, and hardware setups. Experimental results show that \tackle~outperforms the reconstruction-oriented baseline methods on \emph{all} considered settings. We evaluate the proposed end-to-end architecture and training procedure through ablation studies. Our results offer guidance for designing effective task-specific CS-MRI systems in the future. 
    \item We show the generalization of \tackle~to out-of-distribution data by deploying it to a dataset we experimentally acquired using a different acquisition sequence from that of the training data. We further implement a learned 4$\times$-accelerated sequence on a Siemens 3T MRI Skyra scanner. The sequence shortens the scan time from 335 seconds to 84 seconds, a four-fold time reduction as desired, while maintaining high performance. These experiments highlight the real-world practicality of our method.
\end{itemize}

\vspace{-5pt}

\section{Background}
\lblsec{background}

\subsection{Compressed sensing MRI}
\lblsec{csmri_basics}

CS-MRI \cite{lustig2007sparse} refers to accelerating MRI via \emph{compressed sensing (CS)} \cite{candes2006robust}, which aims to reconstruct the underlying image from a set of subsampled $k$-space measurements.

\subsubsection{Basics} The common setup of CS-MRI involves the reconstruction of an image $\xbm \in \C^n$ from its subsampled, noisy $k$-space measurements
\begin{align}
\lbleq{forward_model}
    \ybm := \Mbm \Fbm \xbm + \nbm \in \C^m \quad (m\ll n),
\end{align}
where $\Fbm$ is the Fourier transform, $\Mbm \in \{0,1\}^{m\times n}$ is the subsampling matrix with $\mbm \in \{0,1\}^n$ denoting its subsampling pattern, and $\nbm \in \C^m$ is the complex measurement noise. 
For parallel imaging MRI, the measurements are collected from multiple coils. For the $i$-th coil, the measurements $\ybm_i$ can be expressed as
\begin{align}
\lbleq{forward_model_multicoil}
    \ybm_i := \Mbm \Fbm \Sbm_i \xbm + \nbm_i \in \C^m,
\end{align}
where $\Sbm_i$ is the pixel-wise sensitivity map and $\nbm_i$ is the measurement noise of the $i$-th coil. 
For both settings, we refer to $b:=\|\mbm\|_1$ as the sampling budget and $R := \frac{n}{b}$ as the acceleration ratio of the acquisition. 
Classical CS-MRI enables sampling below the Nyquist-Shannon rate by solving an optimization problem with a regularizer that leverages the structure of MRI images \cite{lustig2008compressed, danielyan2011bm3d, elad2007image, rudin1992nonlinear}. 

\subsubsection{Subsampling patterns}
Subsampling patterns, or masks, in traditional CS-MRI are often generated randomly or handcrafted to have a point spread function (PSF) with low coherence, which leads to better reconstruction performance according to the CS theory.
Popular subsampling patterns include the 2D variable density \cite{lustig2007sparse}, bidirectional Cartesian \cite{wang2009pseudo}, Poisson-disc \cite{vasanawala2011practical}, and continuous-trajectory variable density \cite{chauffert2014variable}, among others \cite{cook1986stochastic, raja2014adaptive}. 
Despite overall effectiveness, these subsampling patterns are designed for generic image reconstruction and not optimized for any specific body part and diagnostic purpose. 
Therefore, these patterns may lead to suboptimal performance for downstream tasks where specific anatomical or pathological information is relevant.

\subsubsection{DL-based reconstruction}
Recently, DL methods have achieved state-of-the-art performance on CS-MRI reconstruction. 
One line of work combines data-driven priors with model-based iterative reconstruction (MBIR) \cite{venkatakrishnan2013plugandplay, romano2016red, ahmad2020plugandplay, kamilov2022plugandplay}. 
Another line of work learns a model-free reconstruction network via end-to-end training \cite{wang2016accelerating, lee2017deep, lee2018deep, yang2017dagan, quan2017compressed}.
A third line of work, known as deep unrolling (DU), combines the characteristics of MBIR and end-to-end training \cite{yang2016deep, hammernik2017learning, zhang2018istanet, aggarwal2019modl, sriram2020end, hosseini2020dense, liu2021sgd, adler2017learned, yaman2020self, gilton2021deep}.
The idea is to ``unroll'' an iterative optimization procedure into a cascade of mappings and train these mappings end-to-end so that they can gradually map a low-resolution input image to a high-quality output reconstruction.
Inheriting the advantage of both MBIR and end-to-end learning, these 
methods exhibit state-of-the-art performance on CS-MRI reconstruction. In this paper, we use a specific kind of unrolled network called E2E-VarNet \cite{sriram2020end} as part of our framework due to its strong performance on the large-scale fastMRI dataset \cite{zbontar2018fastMRI}.

\begin{figure*}[t]
\centering
\includegraphics[width=0.95\textwidth]{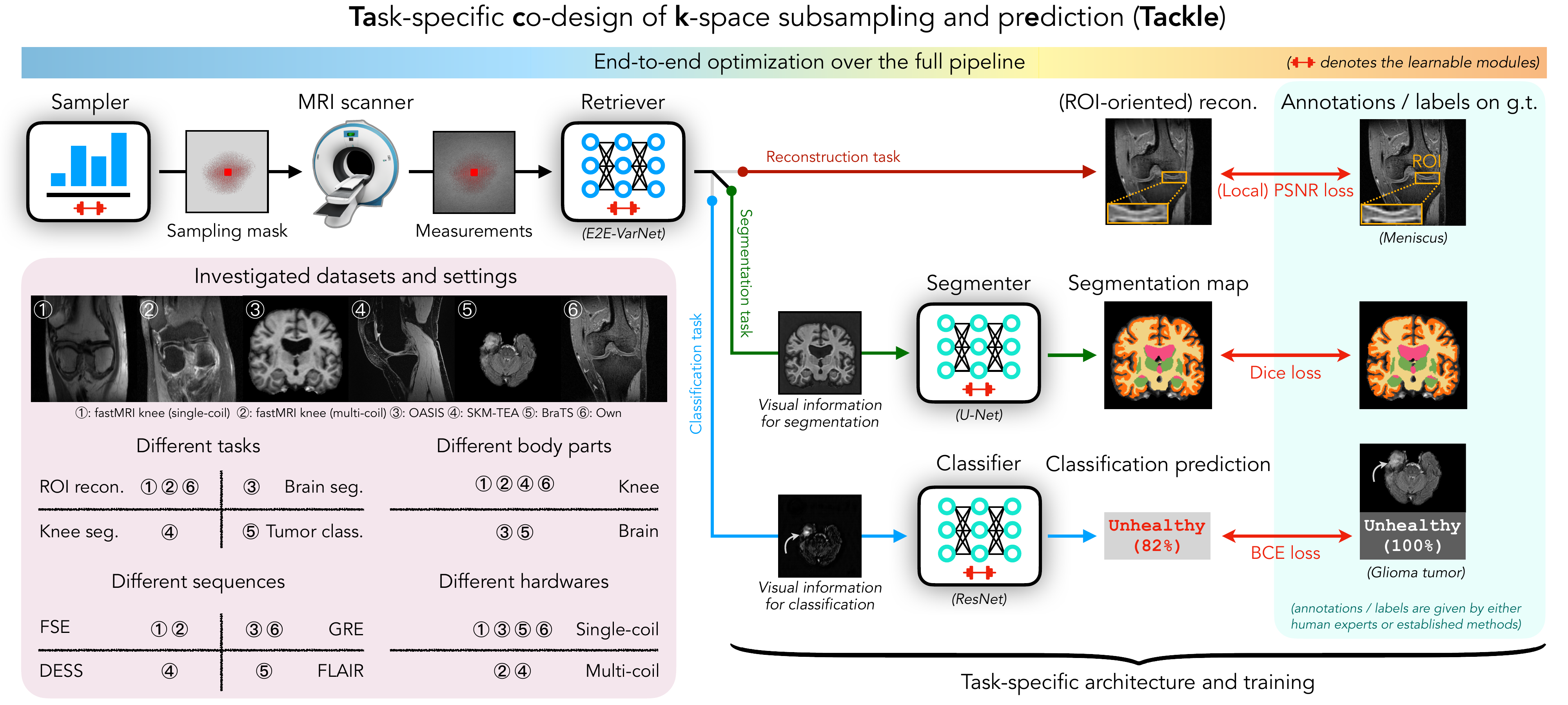}
\caption{
Block diagram of the proposed framework \tackle~and a summary of the investigated datasets and settings.
\tackle~uses a task-specific loss to jointly optimize a sampler, a retriever, and an optional predictor, ranging from scanner-level sampling to human-level diagnosis. 
A summary of the investigated settings is presented in the bottom left panel.
FSE, GRE, DESS, and FLAIR stand for fast spin echo, gradient echo, double-echo steady-state, and fluid-attenuated inversion recovery, respectively.
We comprehensively investigate multiple CS-MRI tasks on a variety of common MRI settings with six datasets. 
}
\vspace{-10pt}
\lblfig{diagram}
\end{figure*}

\vspace{-5pt}

\subsection{Reconstruction-oriented co-design}
\lblsec{learning_design}

The success of DL methods in CS-MRI reconstruction motivates the idea of jointly optimizing acquisition together with reconstruction via end-to-end training. 
Recently, there has been a rapidly growing literature on optimizing a parameterized sampling strategy jointly with a CNN reconstructor \cite{bahadir2019learning, zhang2019reducing, zhang2020extending, aggarwal2020jmodl, alkan2020learning, wang2021bspline, xue20222d, peng2022learning, wang2022stochastic, yang2022fast, martinini2022deep, wang2022joint, zhao2022jojonet, wang2022leaders, chaithya2022hybrid, zibetti2022alternating}. 
These methods have different architectural designs and applicable scenarios, but all rely on the differentiable nature of neural networks to optimize the reconstruction accuracy over the choice of $k$-space measurements. 
The learned subsampling pattern and reconstruction network are thus specific to the dataset.
The end-to-end training enables synergistic cooperation between the learned subsampling pattern and reconstructor, achieving state-of-the-art reconstruction performance.
From a task perspective, however, having a reconstruction is not the end of the workflow.
These methods rely on either human evaluation, a traditional task prediction algorithm, or a CNN for task predictions, which are out of the scope of these papers.

\vspace{-5pt}

\subsection{Task-oriented co-design}
\label{sec:task_oriented}

Recent work has investigated the co-design idea in the context of limited tasks beyond full-FOV reconstruction, such as physical parameter estimation \cite{zou2022joint, xu2022correct, zhao2022jojonet} and segmentation \cite{fan2018segmentation, sun2018joint, weiss2020pilot, wang2021one, wang2024promoting}.
Using task-specific loss functions in their training procedures, these proposed methods demonstrate stronger task performance than methods trained by a reconstruction-only loss. 
Most of these proposed approaches leave either subsampling or prediction as a pre-determined fixed module, and focus on co-designing the other modules \cite{fan2018segmentation, sun2018joint, zou2022joint, xu2022correct, zhao2022jojonet}.
On the other hand, the authors of \cite{weiss2020pilot, wang2021one} proposed to jointly optimize all three steps, and investigated a brain segmentation task using a U-Net reconstructor and predictor.
Although these methods show the potential of extending co-design beyond reconstruction, they are each fine-tuned for one particular task, do not easily accommodate different types of data (e.g., multi-coil), and have not been demonstrated on real out-of-distribution datasets. 
The most relevant work to ours in the literature is a concurrent work by Wang \textit{et al.} \cite{wang2024promoting}, in which the authors presented a thorough investigation of optimizing the entire CS-MRI pipeline for various segmentation problems.
In this work, we cast a wider net for the task-specific CS-MRI co-design problem. 
In particular we demonstrate our unified framework for designing generalized CS-MRI pipelines, \tackle, on three different tasks beyond full FOV reconstruction.
\tackle~performs robustly on this broad range of tasks and experiments, and is implemented and tested on a Siemens scanner.

\vspace{-5pt}

\section{Method}
\lblsec{method}

\reffig{diagram} illustrates the architecture of \tackle.
As a co-design CS-MRI method, \tackle~jointly optimizes the sampler, retriever, and predictor for a task-dependent loss.
In the following subsections, we describe each module in order and more implementation details can be found in Supplement II.

\vspace{-5pt}

\subsection{Sampler}

We consider 2D Cartesian subsampling patterns, i.e. $\mbm \in  \{0,1\}^n$.
We follow \cite{bahadir2019learning, xue20222d, yin2021end} to model the subsampling strategy as the element-wise Bernoulli distribution with a probability vector $\pbm \in [0,1]^n$, i.e. $\mbm_{i} \sim \mathsf{Bern}(\pbm_{i})$. 
To learn the optimal sampling probabilities, we follow the sampler design of \cite{bahadir2019learning}. We optimize a set of parameters $\qbm_i$ that first give us a set of probabilities $\widetilde{\pbm}_i := \mathsf{Sigmoid}(\qbm_i)$. 
We then rescale $\widetilde{\pbm}$ to obtain a probabilistic sampling mask $\pbm$ that would result in $b$ measurements in expectation via Bernoulli sampling:
\begin{align*}
    \pbm = 
    \begin{cases}
      \frac{\alpha}{\beta}\widetilde{\pbm} & \text{if $\beta \geq \alpha$} \\
      \bm{1} - \frac{1-\alpha}{1-\beta}(\bm{1}-\widetilde{\pbm}) & \text{otherwise}
    \end{cases}       
\end{align*}
where $\alpha := \frac{b}{n}$, $\beta := \frac{\|\widetilde{\pbm}\|_1}{n}$, and $\bm{1}$ is the all-one vector. 
During training, the sampler draws a $k$-space sampling mask $\mbm$ by sampling $\mbm_{i} \sim \mathsf{Bern}(\pbm_{i})$. 
We repeatedly sample $\mbm$ until $\|\mbm\|_1 \approx b$ under a small tolerance. 
This sampling process encourages exploration of different patterns and ensures the sampling patterns approximately satisfy the budget constraint. 
Since the sampling process is not differentiable, we use a straight-through estimator to overcome the non-differentiability~\cite{bengio2013estimating}. 
During testing, we set the top $b$ indices of $\pbm$ with the highest probabilities to 1 (to sample) and others to 0 (not to sample). 
This binarization guarantees that the sampling mask strictly satisfies the sampling budget constraint and all slices of a volume share the same sampling mask.
We also allocate $\sfrac{1}{8}$ of the sampling budget for the low-frequency region around the DC component, which we refer to as the pre-select region.
The pre-selected measurements provide auto-calibration signals (ACS) for multi-coil reconstruction and stabilize the training of some baselines. 
Therefore, we include the pre-select region for all experiments for consistency.
More discussion on this can be found in Supplement II.C.
We denote the sampler as $\mathsf{S}_\qbm$ where $\qbm$ is the vector of learnable parameters.

\vspace{-5pt}

\subsection{Retriever}

After acquiring measurements, we employ a retriever to extract visual information from noisy and subsampled $k$-space measurements.
We note that we name the module ``retriever'' instead of ``reconstructor'' because it is jointly optimized with the downstream predictor for non-reconstruction tasks.
Hence, the retriever should not be interpreted as a reconstructor as its output may not be a typical ``reconstruction'' in terms of pixel-wise accuracy.
We denote the retriever as $\mathsf{R}_\thetabm$ where $\thetabm$ is its weights.
We select the E2E-VarNet \cite{sriram2020end} since it is a model-based DU architecture that combines forward model and learning prior, and achieves excellent performance on CS-MRI reconstruction \cite{zbontar2018fastMRI}.
E2E-VarNet also accommodates multi-coil $k$-space data with its ability to estimate coil sensitivity maps.
Specifically, our E2E-VarNet retriever operates in $k$-space and consists of 12 refinement steps, each of which includes a U-Net \cite{ronneberger2015u} with independent weights from each other.
For each U-Net, we use the standard architecture with the following parameters: 2 input and output channels, 18 channels after the first convolution filter, 4 average down-pooling layers, and 4 up-pooling layers.
The final output layer of the retriever is an inverse Fourier transform followed by a root-sum-squares reduction for each pixel over all coils.
The output of the retriever is a batch of magnitude images.
For reconstruction tasks, a loss function will be directly applied to the output.
For non-reconstruction tasks, the output will be fed into an additional predictor module described in the next section.

\vspace{-5pt}

\subsection{Task-specific design: predictor and loss function}
We demonstrate \tackle~on three tasks that together represent a gradual progression from generic full-FOV reconstruction to clinically relevant tasks.

\subsubsection{ROI-oriented reconstruction}
For many MRI scans, only a small region of the FOV is relevant to the reader, so we define a task where we aim to maximize reconstruction quality around that region.
In contrast to the full-FOV reconstruction task, the reconstruction accuracy in this task is only measured over the region-of-interest (ROI) of each image instead of the entire FOV.
We hereafter refer to this task as \textit{ROI-oriented reconstruction}.
This task is a first step from generic full-FOV reconstruction to more specific downstream tasks in CS-MRI. 

There is no predictor for this reconstruction task, and the output of the retriever will directly be used for evaluation.
The evaluation metric we use is the local peak signal-to-noise ratio (PSNR), which is the PSNR within the ROI of an underlying image $\xbm$. Let $\mathcal{R}_{\xbm}$ be the set of indices $i$ that are within the ROI of $\xbm$. Note that $\mathcal{R}_\xbm$ varies from one image $\xbm$ to another.  We define the local PSNR within the ROI as
\begin{align}
\lbleq{local_psnr}
    \mathsf{LocalPSNR}(\xbmhat, \xbm; \mathcal{R}_\xbm) := 10 \log_{10}\frac{\mathsf{max}(\xbm)^2}{\mathsf{LocalMSE}(\xbmhat, \xbm; \mathcal{R}_\xbm)}
\end{align}
where
$\mathsf{LocalMSE}(\xbmhat, \xbm; \mathcal{R}_\xbm) := \frac{1}{|\mathcal{R}_\xbm|} \sum_{i \in \mathcal{R}_\xbm} (\xbmhat_{i} - \xbm_{i})^2$
and $\mathsf{max}(\xbm)$ is the largest pixel value of $\xbm$. 
We optimize our model for the local reconstruction quality using $\mathcal{L}_\text{ROI}(\xbmhat, \xbm) := -\mathsf{Local PSNR}(\xbmhat, \xbm; \mathcal{R}_\xbm)$ as the training loss. 

\subsubsection{Tissue segmentation} 
For this task, we aim to predict segmentation maps of different body tissues. 
Accurately segmenting a tissue from the rest of the organ provides important anatomical and pathological information~\cite{heimann2010segmentation, schick2016tissue, raj2018automatic}.
Conventional segmentation workflow involves human evaluation and traditional algorithms, which often require standard reconstructions of certain contrasts as input \cite{fischl2012freesurfer}.
On the contrary, \tackle~does not require reconstruction as a necessary intermediate step, and is optimized for segmentation performance in an end-to-end fashion.

We include an additional predictor $\mathsf{P}_\phibm$ with weights $\phibm$ subsequent to the retriever.
We choose the U-Net architecture due to its ability of solving medical image analysis tasks~\cite{ronneberger2015u, balakrishnan2019voxelmorph, ghodrati2019mr}. 
The specific parameters are: 1 input channel, $c$ output channels (where $c$ is the number of segmentation classes), 64 channels after the first convolution filter, 4 average down-pooling layers, and 4 up-pooling layers.

We use the Dice score \cite{dice1945measures, zou2004statistical, milletari2016v} as the evaluation metric. 
The Dice score measures the degree of overlap between two segmentation maps and takes a value between 0 (no overlap) and 1 (perfect overlap). During training, we employ the Dice loss $\mathcal{L}_\text{seg.}(\zbmhat, \zbm) := 1 - \mathsf{DiceScore}(\zbmhat, \zbm)$.
For both training and evaluation, we apply a Softmax function across all the classes for each pixel and then calculate the Dice loss/score.
During the evaluation, we apply an additional binarization step where we set the class with highest value after Softmax as 1 and others as 0. 
In this way, we assign each pixel of the predicted segmentation map $\zbmhat$ to exactly one class.

\subsubsection{Pathology classification}
\lblsec{pathology_classification}
The third task we consider is to determine whether a potential pathology exists in an MRI image, such as a suspected tumor. 
Using algorithms to automatically analyze MRI scans could lead to improved diagnosis accuracy in clinical practice \cite{menze2014multimodal}. 
We formulate this task as a binary image classification problem, where the negative class means the underlying image $\xbm$ does not contain any pathology lesion and the positive class means it does contain a lesion.
Through this proof-of-concept classification task, we go beyond pixel-level problems and show the benefit of task-specific co-design for solving an image-level problem.

Similar to the segmentation task, we include an additional predictor in the pipeline, which we also denote as $\mathsf{P}_\phibm$ to simplify notations.
Specifically, we choose the ResNet \cite{he2016deep}, which is an established architecture for computer vision tasks, especially image classification. 
We use the standard ResNet18 architecture except for using 1 input channel and 2 output dimensions.

We use the binary cross entropy (BCE) as the loss function for this classification task, $\mathcal{L}_\text{class.}(\zbmhat, \zbm) := \mathsf{BCE}(\zbmhat, \zbm)$.
For evaluation metrics, we consider both the classification accuracy ($\mathsf{ClsAcc} := \frac{\text{TP}+\text{TN}}{\text{TP}+\text{TN}+\text{FP}+\text{FN}}$) and the $F_1$ score ($\mathsf{F_1 \, score} := \frac{2\text{TP}}{2\text{TP}+\text{FP}+\text{FN}}$) where TP, TN, FP, and FN are the number of True Positive, True Negative, False Positive, and False Negative, respectively. The classification accuracy is more interpretable, while the $F_1$ score is more robust to class imbalance. So we include both metrics for a more comprehensive evaluation.

\vspace{-5pt}

\subsection{Training procedure}
\lblsec{training_procedure}
We summarize the training objective for each task as follows:
\begin{itemize}
    \item ROI-oriented reconstruction: 
    $$\min_{\qbm, \thetabm} \mathcal{L}_\text{ROI}\left(\mathsf{R}_\thetabm \left( \mathsf{S}_\qbm \odot \kbm \right), \xbm \right)$$
    \item Segmentation or classification: 
    $$\min_{\qbm, \thetabm, \phibm} \mathcal{L}_\text{seg. / class.}\left(\mathsf{P}_\phibm \left( \mathsf{R}_\thetabm \left( \mathsf{S}_\qbm \odot \kbm \right) \right), \zbm \right)$$
\end{itemize}
where $\kbm \in \C^n$ contains all $k$-space measurements of $\xbm$ and $\odot$ denotes element-wise multiplication. 

When performing end-to-end training over multiple stages, we empirically observed that a model trained from scratch tends to run into either optimization (hard to train) or generalization (unable to generalize) issues.
Some prior works address these problems using a hybrid of reconstruction and task-dependent loss \cite{sun2018joint, weiss2020pilot, wang2021one, xu2022correct, zou2022joint}.
This approach requires tuning a weight parameter that balances the two losses.
We adopt an alternative approach that avoids tuning this additional parameter.
Specifically, we first train the sampler and retriever jointly with a full-FOV PSNR loss until convergence:
$$\min_{\qbm, \thetabm} \mathcal{L}_\text{FOV}\left(\mathsf{R}_\thetabm \left( \mathsf{S}_\qbm \odot \kbm \right), \xbm \right)$$
where $\mathcal{L}_\text{FOV}(\xbmhat, \xbm) := -\mathsf{PSNR}(\xbmhat, \xbm)$.
We refer to this as the pre-training step in later sections.
With the weights learned for the sampler and retriever, we then add the predictor (initialized with random weights) into the framework and fine-tune all three components.
We find that the pre-training step allows the model to better learn task-specific strategies, as demonstrated by an ablation study in \refsec{ablation_architecture}.
This training procedure mimics the training of foundation models in state-of-the-art language and vision models, which are first pre-trained on a general task and then fine-tuned for more specific tasks.
Similar procedures can be found in other task-specific co-design papers, such as \cite{sun2018joint, fan2018segmentation}.

\section{Experiments on Large-Scale Datasets}
\lblsec{experiments}
We first demonstrate the effectiveness of our framework on the three considered tasks using large-scale datasets. 
We categorize all the investigated datasets and settings in the bottom left panel of \reffig{diagram}.
For each task, we demonstrate that the proposed task-specific co-design framework achieves better performance than baselines that separately design reconstruction and prediction.
We abbreviate different variants of the proposed method and baselines in the following way based on their task and training procedure:

\begin{table}[ht]
\newcolumntype{C}{>{\centering\arraybackslash}p{34pt}}
\newcolumntype{D}{>{\centering\arraybackslash}p{128pt}}
\newcolumntype{E}{>{\centering\arraybackslash}p{57pt}}
\centering
\begin{tabularx}{254pt}{CDE}
\toprule 
Task & Training procedure and loss & Notation example \\ \midrule
\multirow{2}{*}{(ROI) recon.} & $\mathsf{S}$\&$\mathsf{R}$: PSNR loss & LP+UN$_\text{FOV}$  \\
 & $\mathsf{S}$\&$\mathsf{R}$: local PSNR loss (w/ pre-training) & \tackleroi  \\ \midrule
\multirow{2}{*}{Tissue seg.} & $\mathsf{S}$\&$\mathsf{R}$: PSNR loss $\to$ $\mathsf{P}$: Dice loss & PD+UN$_\text{recon.}$  \\
 & $\mathsf{S}$\&$\mathsf{R}$\&$\mathsf{P}$: Dice loss (w/ pre-training) & \tackleseg  \\
 \midrule
\multirow{2}{*}{Patho. class.} & $\mathsf{S}$\&$\mathsf{R}$: PSNR loss $\to$ $\mathsf{P}$: BCE loss & \louperecon  \\
 & $\mathsf{S}$\&$\mathsf{R}$\&$\mathsf{P}$: BCE loss (w/ pre-training) & \tackleclass  \\
\bottomrule
\multicolumn{3}{l}{$\mathsf{S}$: sampler, $\mathsf{R}$: retriever, $\mathsf{P}$: predictor} \\
\multicolumn{3}{l}{$\to$: separate training stages for reconstruction and prediction}
\vspace{-5pt}
\end{tabularx}
\end{table}
\noindent To clarify, the subscript ``recon.'' for the segmentation and classification methods means that the sampler and retriever are trained for full-FOV reconstruction, and a predictor is subsequently trained for the downstream task with the sampler and retriever fixed.
This is equivalent to training a predictor with the reconstructed images by these methods as input for the downstream task.

\begin{figure*}[t]
\centering
\includegraphics[width=0.95\textwidth]{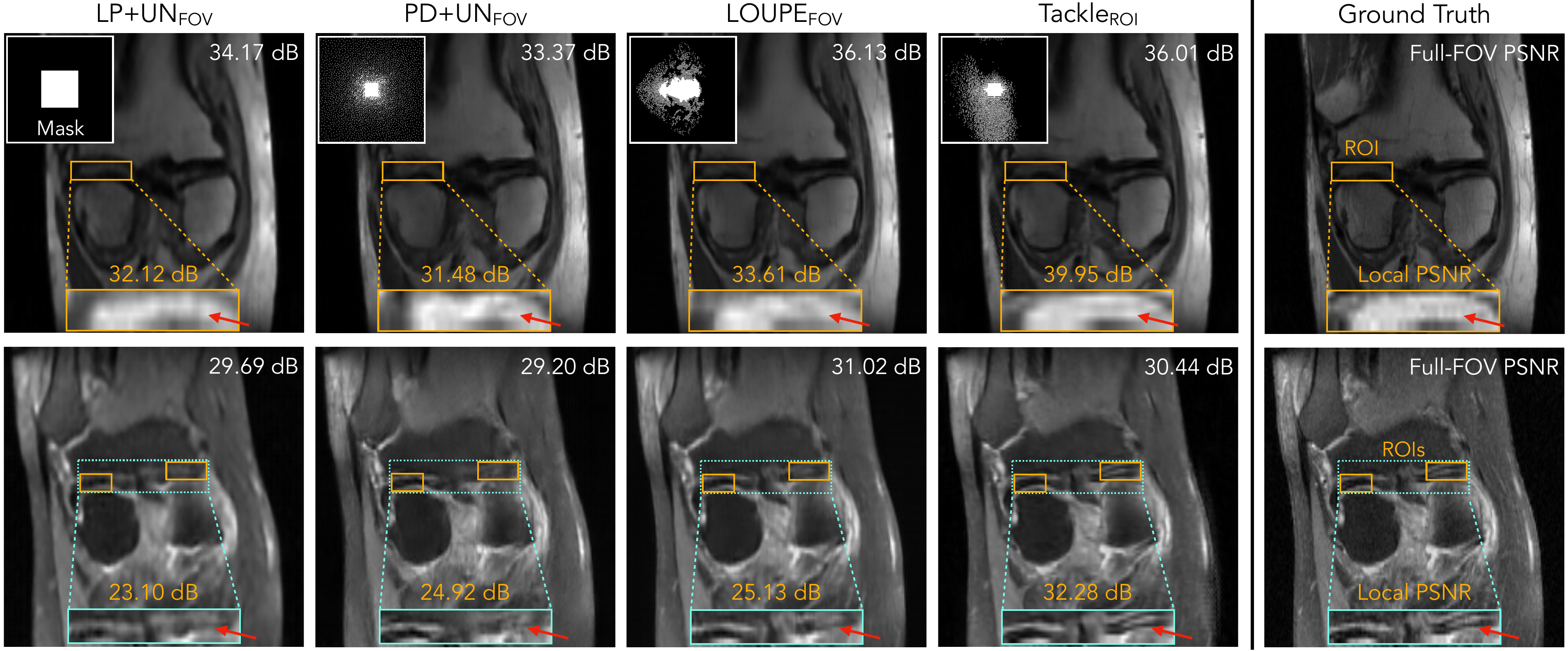}
\caption{
Visual examples of two Meniscus Tear samples reconstructed by different methods in the 16$\times$ acceleration single-coil setting. 
For each reconstruction, the full-FOV PSNR is labeled in white, and the local PSNR for the ROI is in orange. 
Note how \tackleroi~recovers the structure and details of the ROI more accurately than the two baselines, as indicated by the red arrows.
The better recovery of \tackleroi~over the ROI leads to a more accurate diagnosis of the Meniscus Tear.
We emphasize that the location of the ROI is not an input to any of these models and is only used for evaluating the accuracy of each method on the region that contains the pathology.
}
\vspace{-10pt}
\lblfig{local_recon}
\end{figure*}

\begin{table}[t]
\centering
\scriptsize
\caption{Comparison of average test local peak signal-to-noise ratio (Local PSNR) in decibel (dB) within Meniscus Tear ROIs  under different acceleration ratios ($R$)}
\begin{tabular}{cccccc}
\toprule 
Data & $R$ & LP+UN$_\text{FOV}$ & PD+UN$_\text{FOV}$ & \loupefov & \tackleroi \\ \midrule
\multirow{2}{*}{Single-coil} & 8 & 26.95 & 28.23 & 30.32 & \textbf{34.04} \\
 & 16 & 25.16 & 26.05 & 27.32 & \textbf{31.54} \\ \midrule
\multirow{2}{*}{Multi-coil} & 8 & 27.55 & 32.68 & 34.88 & \textbf{40.65} \\
 & 16 & 26.02 & 30.00 & 31.79 & \textbf{37.89} \\
\bottomrule 
\end{tabular}
\lbltab{exp_roi_mt}
\vspace{-10pt}
\end{table}

\begin{figure}[t]
\centering
\includegraphics[width=0.48\textwidth]{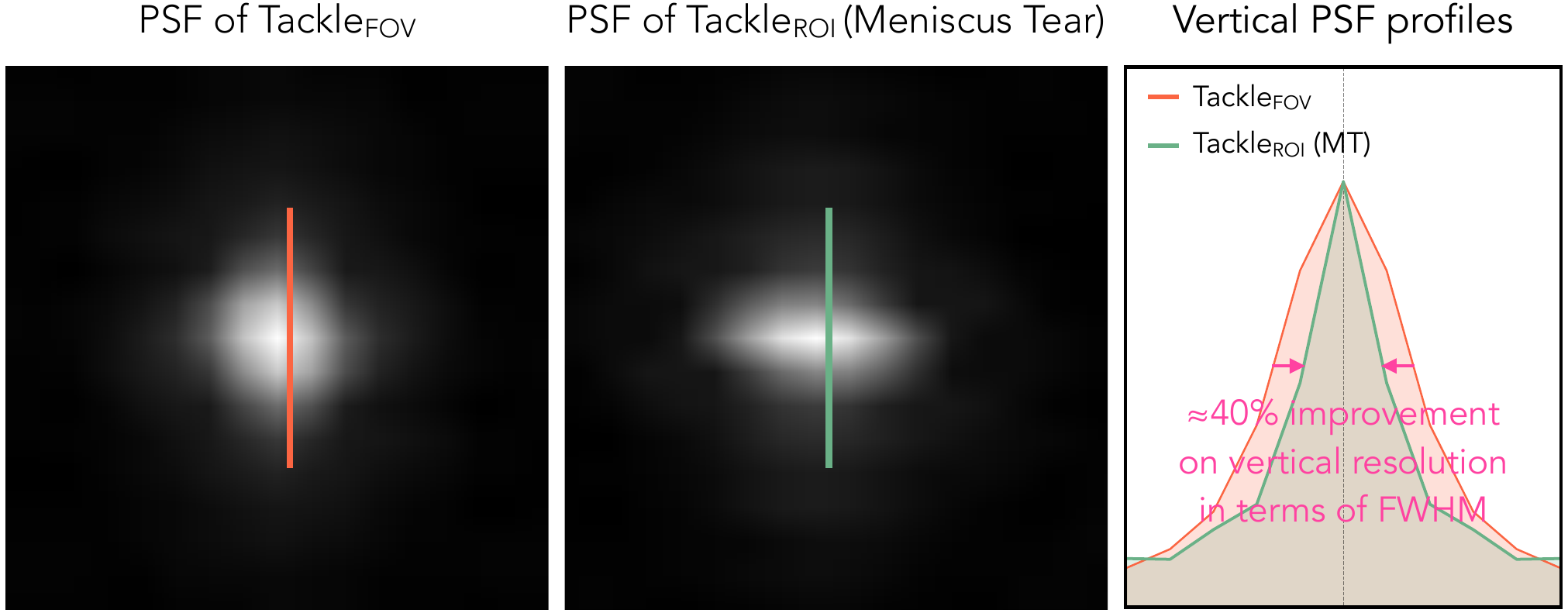}
\caption{
Comparison of a subsampling PSF optimized for full-FOV reconstruction and another optimized for the reconstruction of menicus tear (MT) ROIs. 
Optimizing for MT ROI reconstruction leads to around 40\% improvement on the vertical resolution in terms of the \emph{full width at half maximum (FWHM)}, as shown by the PSF profiles in the bottom panel.
This improved vertical resolution leads to a better reconstruction of the meniscus that has horizontal anatomy.
}
\vspace{-10pt}
\lblfig{psf}
\end{figure}

\vspace{-5pt}

\subsection{ROI-oriented reconstruction}
\lblsec{exp_roi}

\noindent \textbf{Dataset and setup} \quad
For the ROI-oriented reconstruction task, we use the images and raw single- and multi-coil $k$-space data from the fastMRI+ knee dataset~\cite{zbontar2018fastMRI, zhao2022fastmri+}, which contains bounding box annotations for knee pathologies.
Specifically, we investigate the most common knee pathology in the dataset called ``Meniscus Tear'' (MT).
Each image $\xbm$ in the dataset contains at least one rectangular bounding box annotation $\mathcal{R}_\xbm$, which is drawn to include all the pathology but exclude the normal surrounding anatomy \cite{zhao2022fastmri+}. 
Therefore the local image quality within each bounding box (i.e. ROI) is more indicative of the quality for pathology assessment than a metric over the entire FOV.
We emphasize that the location of the bounding box $\mathcal{R}_\xbm$ \emph{varies sample by sample} and is \emph{never} an input to any method during inference.
$\mathcal{R}_\xbm$ is only used for calculating the training loss and evaluating the local PSNR during test time according to \refeq{local_psnr}. 
Hence, the local PSNR performance reflects the quality of reconstructions by different methods for assessing the considered pathological lesions in the ROIs.

\noindent \textbf{Baselines} \quad 
We compare \tackleroi~with three full-FOV reconstruction-oriented baselines.
\begin{itemize}
    \item \textit{\loupefov:} Proposed in \cite{bahadir2019learning}, \loupefov~jointly optimizes a sampler and a residual U-Net reconstructor.
    \item \textit{Low-pass + U-Net$_\text{FOV}$ (LP+UN$_\text{FOV}$):} substitute the sampler in \loupefov~with a fixed low-pass filter sampling pattern.
    \item \textit{Poisson-disc + U-Net$_\text{FOV}$ (PD+UN$_\text{FOV}$):} substitute the sampler in \loupefov~with a Poisson-disc sampling pattern drawn from a variable density distribution and generated with the $\texttt{sigpy.mri.poisson}$ function in the \texttt{SigPy} package\footnote{\href{https://github.com/mikgroup/sigpy}{https://github.com/mikgroup/sigpy}}.
\end{itemize}

\noindent \textbf{Results} \quad
We compare the average local PSNR of our method and other baselines over the test set in \reftab{exp_roi_mt}. For all settings, \tackle~outperforms other baselines designed for full-FOV reconstruction by at least 3 dB, indicating a significant improvement of image quality within the ROI. 

In \reffig{local_recon}, we provide example reconstructions by our method and three baseline methods. 
For each reconstruction, a zoom-in on its ROI is provided on the bottom with the corresponding local PSNR value labeled above in orange, and its full-FOV PSNR is labeled on the top right corner. 
As shown in the ground truth of the MT example, a meniscus tear is indicated by a streak (dark in the top row and bright in the bottom row) that is present on the meniscus (bright in the top row and dark in the bottom row), as indicated by the red pointers. 
To accurately detect the existence and assess the severity of a meniscus tear, a reconstruction should clearly show the boundaries of the meniscus and details of the tear. 
However, the ROIs of both LP+UN$_\text{FOV}$ and \loupefov~reconstructions contain significant reconstruction artifacts that disguise the tear (see the red arrows). 
On the other hand, \tackleroi~preserves the details of the tear and contains fewer artifacts than the baselines, providing a more accurate ROI reconstruction with a higher diagnostic value. 

In Supplement IV we also include a validation of \tackleroi~on images that either are healthy or contain pathologies other than the meniscus tear. 
Although \tackleroi~is not designed to generalize across different pathologies, we empirically find that \tackleroi~still yields high-fidelity reconstructions for out-of-distribution images so that the pathologies on these images remain detectable. 
We also find that \tackleroi~generalizes consistently across the three acceleration ratios (4$\times$, 8$\times$, and 16$\times$) for the fastMRI+ dataset.

\noindent \textbf{Discussion} \quad
Enhancing local ROIs for MRI may seem counter-intuitive, because the acquisition happens in $k$-space; each frequency measurement in theory corresponds to the entire FOV.
Here we understand the feasibility via a PSF analysis. 
Consider the zero-filled reconstruction $\xbmtilde$ from some (noiseless) single-coil $k$-space data:
$$\xbmtilde := \Fbm^{-1} \left(\mbm \odot \left(\Fbm \xbm \right)\right) = \left(\Fbm^{-1}\mbm\right) * \xbm$$
where $*$ denotes convolution and the second equality holds due to the Fourier convolution theorem. 
Here, $\Fbm^{-1}\mbm$ is the PSF of the subsampling mask $\mbm$ and determines the resolution of the CS-MRI system. 
We visualize the PSF of a sampling mask trained for full-FOV reconstruction and another trained for MT ROIs reconstruction with the same sampling budget in \reffig{psf}. 
We plot the PSF profiles in the vertical direction around the main lobes. 
The PSF learned for MT ROI reconstruction has around 40\% improvement in vertical resolution in terms of \emph{full width at half maximum (FWHM)} of the PSF profiles.
Since MT ROIs contains the thin horizontal anatomy of the meniscus, it makes sense that the learned subsampling pattern has a narrower PSF profile (and thus higher resolution) in the vertical direction.
This comparison demonstrates that the improvement on ROIs is partly due to the capability of our model to optimize the subsampling PSF for local ROI anatomy via co-design.
This is particularly beneficial when there is a mismatch between the optimal subsampling PSF for full-FOV reconstruction and that for ROI reconstruction due to directional anatomical structure, which is the case for MT ROI reconstruction.

\begin{table}[t]
\centering
\scriptsize
\caption{Comparison of average test Dice score on the SKM-TEA dataset \cite{desai2021skmtea} for segmenting four knee tissues under different acceleration ratios ($R$)}
\begin{tabular}{cccccc}
\toprule 
$R$ & PD+UN$_\text{recon.}$ & \louperecon & SemuNet & \tacklerecon & \tackleseg \\ \midrule
16 & 0.7843 & 0.7888 & 0.8108 & 0.8232 & \textbf{0.8532} \\
64 & 0.7486 & 0.6715 & 0.7741 & 0.8145 & \textbf{0.8357} \\
\bottomrule
\end{tabular}
\vspace{-10pt}
\lbltab{exp_skmtea}
\end{table}

\begin{figure}[t]
\centering
\includegraphics[width=0.45\textwidth]{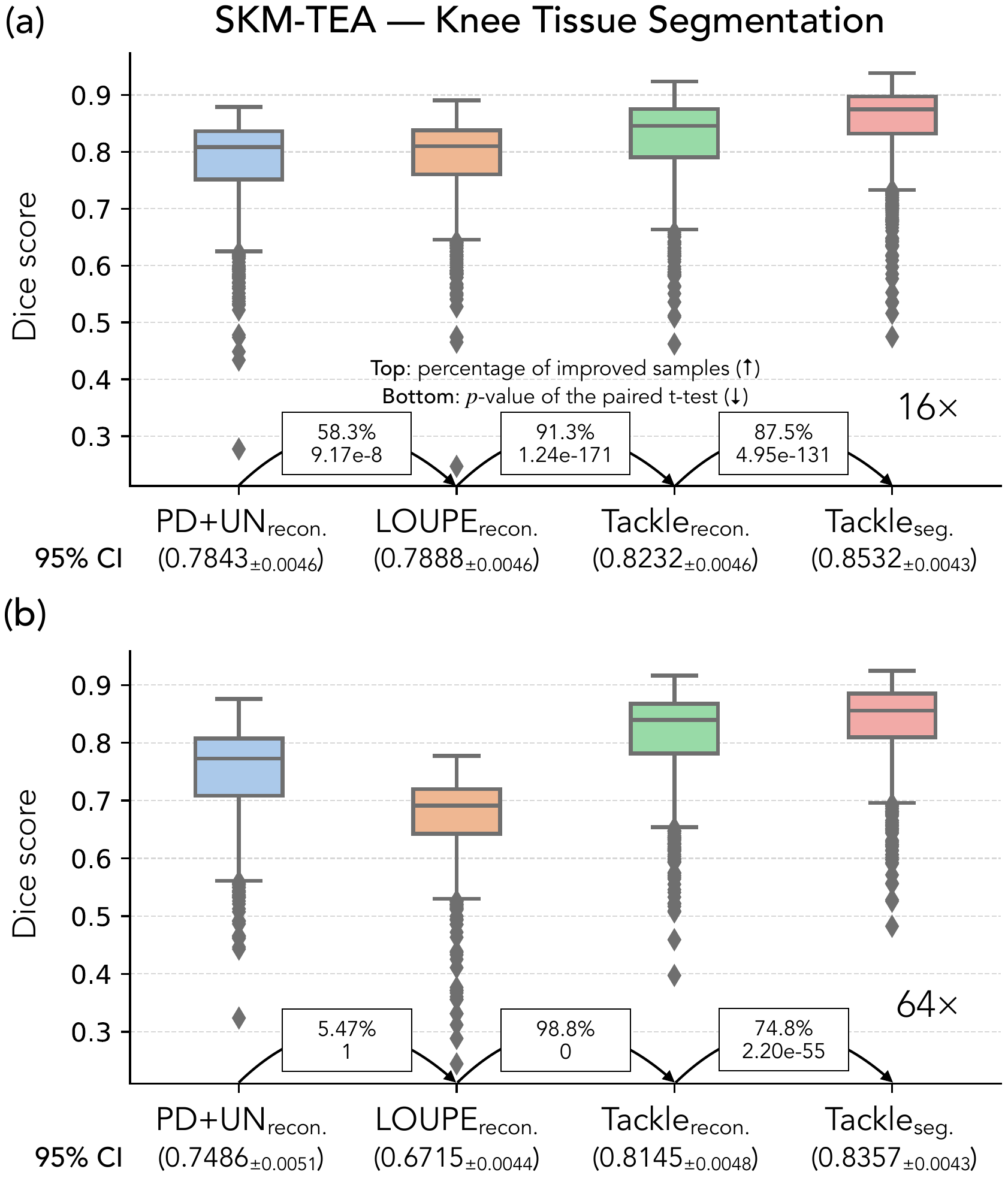}
\caption{
Box plots of the knee tissue segmentation results under 16$\times$ (a) and 64$\times$ (b).
Within the rectangle between each pair of methods, the top number is the percentage of samples that get improved and the bottom number is the $p$-value given by the paired samples t-test.
A higher percentage and lower $p$-value indicate a more significant improvement.
We also provide the 95\% confidence intervals for all methods below their names.
For both acceleration ratios, \tackleseg~outperforms other baselines in terms of all the statistical measures.
}
\lblfig{knee_box}
\vspace{-10pt}
\end{figure}

\begin{figure*}[t]
\centering
\includegraphics[width=0.95\textwidth]{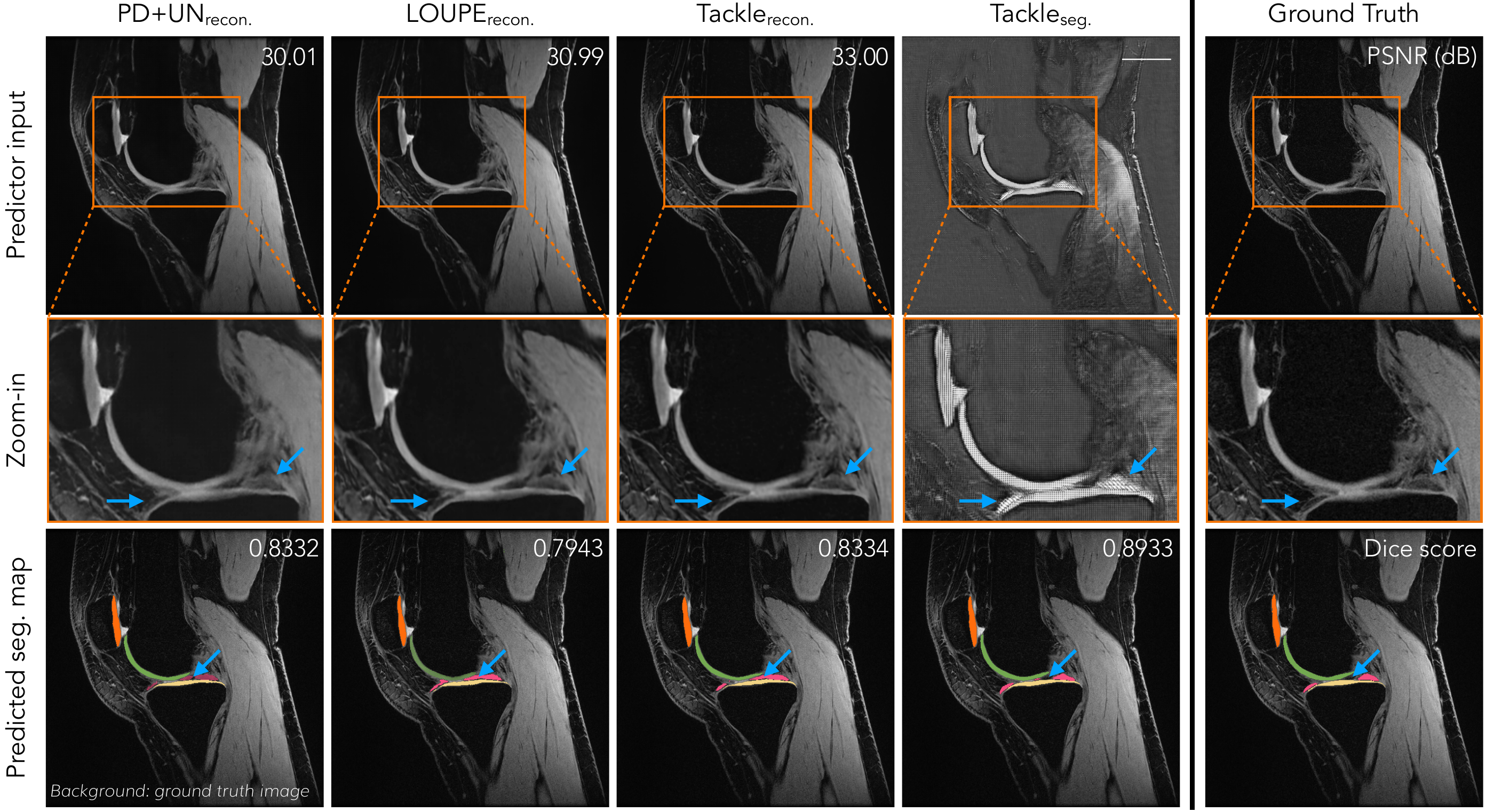}
\caption{
Comparison of segmentation results under 16$\times$ acceleration on one sample of the SKM-TEA dataset. 
We show the input of the predictor in the first row, a zoom-in on the region that contains the tissues to be segmented in the second row, and the output of the predictor in the third row.
Note that \tackleseg~circumvents the typical ``reconstruction'' in terms of pixel-wise similarity with the ground truth image.
Instead, it learns a feature map that accurately localizes the anatomy, leading to better segmentation prediction than other baselines both for this sample and on average over the test set (\reftab{exp_skmtea}).
}
\lblfig{knee_examples}
\vspace{-10pt}
\end{figure*}

\vspace{-5pt}

\subsection{Knee tissue segmentation\protect\footnote{See brain segmentation results in Supplement I.}}
\lblsec{exp_seg}

\noindent \textbf{Dataset and setup} \quad
This study involves segmenting four types of knee tissues: the patellar cartilage, the femoral cartilage, the tibial cartilage, and the meniscus.
We use the \textit{Stanford Knee MRI with Multi-Task Evaluation (SKM-TEA)} dataset \cite{desai2021skmtea}, which contains pixel-level segmentation maps of the four tissues. 
Specifically, we use the raw 3D multi-coil $k$-space measurements of knee images and take 1D inverse Fourier transform along the left-to-right direction to obtain 2D $k$-space of sagittal slices. 
We train each method to minimize the Dice loss until convergence and select the model with the highest Dice score on the validation set.

\noindent \textbf{Baselines} \quad
We compare \tackle$_\text{seg.}$~with four baselines. 
\begin{itemize}
    \item \textit{\louperecon:} \louperecon~is a baseline based on \loupefov. We first train a \loupefov~model for the full-FOV reconstruction task and then use the reconstructed images to separately train a segmentation network.
    \item \textit{Poisson-disc + U-Net$_\text{recon.}$ (PD+UN$_\text{recon.}$):} same as \louperecon~except that the sampler is fixed to be a Poisson-disc sampling mask.
    \item \textit{\tacklerecon:} same as \louperecon~except for using the proposed architecture of \tackle.
    \item \textit{SemuNet:} Proposed in \cite{wang2021one}, SemuNet uses a hybrid of $\ell_1$ reconstruction loss and cross-entropy segmentation loss.
\end{itemize}

\noindent \textbf{Results} \quad
We provide a quantitative comparison in \reftab{exp_skmtea} and a boxplot comparison in \reffig{knee_box}.
Within the rectangle between each pair of methods in \reffig{knee_box}, the top number is the percentage of samples that get improved and the bottom number is the $p$-value given by the paired samples t-test.
With an improved architecture, \tacklerecon~already outperforms the other baselines.
Nevertheless, the segmentation-oriented method \tackleseg~achieves even better performance on both 16$\times$ and 64$\times$ accelerations.
\tackleseg~also significantly outperforms SemuNet on both acceleration ratios and has a much smaller performance drop from 16$\times$ to 64$\times$ than SemuNet, indicating that the proposed approach is more robust to high acceleration ratios. 
We further provide some visual examples in \reffig{knee_examples}. 
The first row visualizes the input of the predictor by different methods, where each image is labelled by its PSNR value on the top right corner.
The last row shows the predicted segmentation maps by different methods, where each prediction is labelled by its Dice score on the top right corner.
The blue arrows point out the locations where \tackleseg~provides more accurate reconstructions than other reconstruction-oriented baselines. 
We also provide a zoom-in on the region that contains the segmented tissues in the second row. 

\noindent \textbf{Discussion} \quad
We note that \tackleseg~learns an intermediate feature map as the input to the predictor, which circumvents a typically ``good'' reconstruction; 
it is interesting how the retriever produces an image where different knee tissues to be segmented have distinctive textures, which are easy to distinguish both from the background and from each other.
Even though this feature map is not a typical ``reconstruction'' in terms of pixel-wise accuracy, it still accurately localizes the anatomy of the tissues to be segmented.
We highlight that \tacklerecon~provides a high-fidelity reconstruction of the entire FOV with a PSNR of 33.00 dB, which demonstrates that our model is well capable of doing the full-FOV reconstruction task accurately. 
However, \tackleseg~still outperforms \tacklerecon~in terms of segmentation performance in \reffig{knee_examples} and on average over the dataset in \reftab{ablation_codesign} (see \refsec{ablation_codesign} for more details).
This observation demonstrates that finding the most accurate full-FOV reconstruction does not necessarily lead to the optimal result on the considered segmentation task.

\vspace{-5pt}

\subsection{Pathology (tumor) classification}
\lblsec{exp_class}

\noindent \textbf{Dataset and setup} \quad
In this section, we demonstrate the effectiveness of the proposed method at detecting the existence of gliomas, a common type of brain tumors in adults.
We use the images acquired by the FLAIR sequence in the Multimodal Brain Tumor Image Segmentation Benchmark (BRATS) dataset \cite{menze2014multimodal}. 
To obtain an image-level label of the existence of a tumor, we aggregate the pixel-level peritumoral edema (ED) segmentation annotations in the BRATS dataset by checking whether there exists any positive pixel in the segmentation map: negative (healthy) means there is no ED pixel, while positive (unhealthy) means there is at least one ED pixel.
We simulate the single-coil $k$-space data for each image by taking the Fourier transform of the image and adding complex additive white Gaussian noise (AWGN), according to the forward model in \refeq{forward_model}. 
The standard deviation of the noise for each image is 0.05\% of the magnitude of the DC component. 
We train all models using the BCE loss and evaluate them using the classification accuracy and $F_1$ score as described in \refsec{pathology_classification}.

\noindent \textbf{Baselines} \quad
We compare the proposed method \tackleclass~with the first three baselines as in \refsec{exp_seg} except that the predictor of each baseline is subsequently trained for pathology classification rather than tissue segmentation (with input images optimized for full-FOV reconstruction). 
We do not include SemuNet here because it is originally proposed for the segmentation task only.

\begin{table}[t]
\centering
\scriptsize
\caption{Comparison of average test accuracy on the pathology classification task under different acceleration ratios ($R$)}
\begin{tabular}{cccccc}
\toprule 
Metric & $R$ & PD+UN$_\text{recon.}$ & \louperecon & \tacklerecon & \tackleclass \\ \midrule
\multirow{2}{*}{Cls. acc.} & 16 & 0.9016 & 0.9024 & 0.9062 & \textbf{0.9159} \\
 & 64 & 0.8809 & 0.8930 & 0.9054 & \textbf{0.9136} \\ \midrule
\multirow{2}{*}{$F_1$ score} & 16 & 0.8853 & 0.8846 & 0.8929 & \textbf{0.9039} \\
 & 64 & 0.8628 & 0.8768 & 0.8910 & \textbf{0.8992} \\
\bottomrule
\end{tabular}
\vspace{-10pt}
\lbltab{exp_class}
\end{table}

\noindent \textbf{Results} \quad
In \reftab{exp_class}, we compare the classification-oriented method, \tackleclass, with reconstruction-oriented baselines, and find that \tackleclass~achieves higher classification accuracy under both performance metrics.
Specifically, \tackleclass~outperforms the existing reconstruction-oriented baseline \louperecon~by around 2\% in the extreme 64$\times$ accelerated acquisition scenario. 
Both variants of \tackle~maintain competitive performance under the highly accelerated setting ($R$=64), while PD+UN$_\text{recon.}$ and \louperecon~suffer from significant performance degradation. 
Note that \tackleclass~outperforms \tacklerecon~by more than 0.8\% in both cases, despite having the same architecture. 
We also visualize and compare the classification performance of \tackleclass~and \louperecon~under 16$\times$ acceleration in \reffig{class_confusion_matrices}, using confusion matrices. 
The results show that \tackleclass~has substantially fewer false negatives (bottom left) and a higher overall accuracy compared to \louperecon. 

\begin{figure}[t]
\centering
\includegraphics[width=0.45\textwidth]{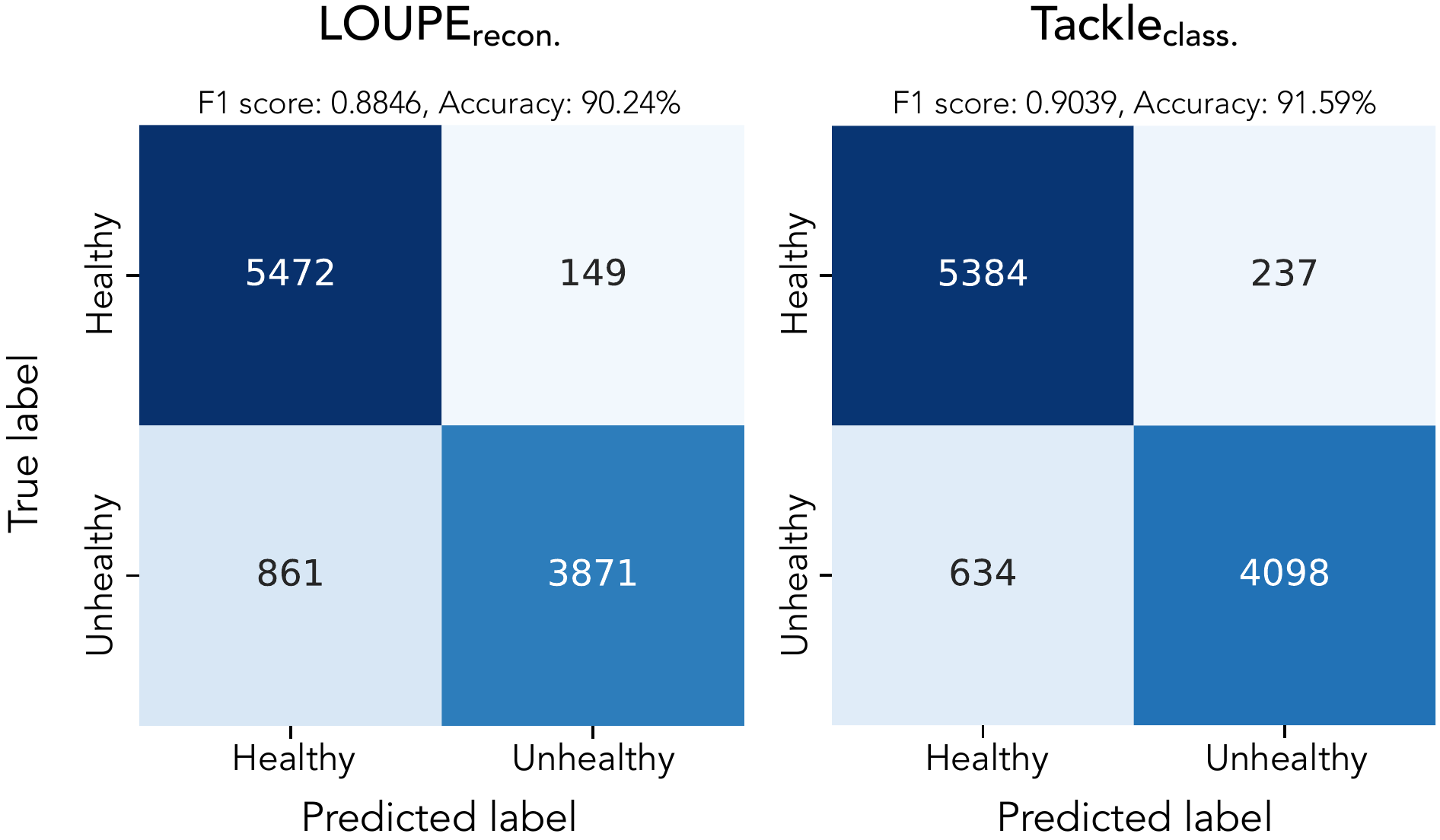}
\caption{
Confusion matrices of the classification results by \louperecon~and \tackleclass. 
Overall, \tackleclass~achieves higher accuracy in terms of both classification accuracy and $F_1$ score than \louperecon. \tackleclass~also has a significantly lower number of false negatives (bottom left) compared to \louperecon, which could lead to more patients receiving early treatment.
}
\lblfig{class_confusion_matrices}
\vspace{-10pt}
\end{figure}

\section{Validation on an Experimentally Collected Out-of-Distribution Dataset}
\lblsec{martinos}

In practice, creating a large well-annotated training set for a specific task can be very time-consuming or even infeasible.
To demonstrate the immediate benefit of our method in a real-world setting, we conduct a validation of \tackle~on the ROI-oriented reconstruction task using experimentally collected data that is out of the distribution of the training data. 
Specifically, we train a \tackle~model on a large-scale dataset (fastMRI in this case) and directly test it on raw $k$-space data collected by \textit{different hardware using a different type of sequence} from that of the training. 
Even without extra fine-tuning or test-time optimization, the learned ROI-specific model provides improved reconstructions on meniscus ROIs.
In the following subsections, we present the details of this experiment.

\noindent \textbf{Data acquisition and processing} \quad
Two subjects were scanned at the Massachusetts General Hospital in accordance with institutional review board guidelines.
Their right knees were scanned by a 3D-encoded Cartesian gradient-echo sequence with a 3 Tesla MRI scanner (Model: Skyra; Siemens Healthcare, Erlangen, Germany) and a single-channel extremity coil.
To implement the 2D subsampling pattern in the coronal plane, we used a transversal orientation with the frequency encoding direction ($k_x$) pointing into the knee cap (anterior-posterior), so that the two phase encoding directions were left-right ($k_y$) and superior-inferior ($k_z$), respectively.
The acquisition parameters were as follows: TE/TR=4.8/9.1ms, FOV=192$\times$192$\times$192mm$^3$, resolution=1$\times$1$\times$1mm$^3$, flip angle=10$^\circ$. 
The total acquisition time of obtaining the fully sampled data for each subject was 5 minutes and 35 seconds. 
The raw $k$-space data had the shape of 192$\times$192$\times$192 ($k_x \times k_y \times k_z$). 
We applied the 1D inverse Fourier transform along $k_x$ for downstream processing. 
Specifically, we took the middle 40 slices of each volume and annotated bounding boxes around the meniscus region using an image labelling tool\footnote{\href{https://github.com/heartexlabs/labelImg}{https://github.com/heartexlabs/labelImg}}. 
Efforts were made such that the locations and sizes of the bounding boxes roughly match those in the fastMRI MT dataset. 
We emphasize that these bounding boxes are \textit{only} for the purpose of measuring the accuracy of different models on reconstructing the meniscus region.
The locations of the annotated ROIs are \textit{not} the input to any of the tested models.

\begin{figure}[t]
\centering
\includegraphics[width=0.45\textwidth]{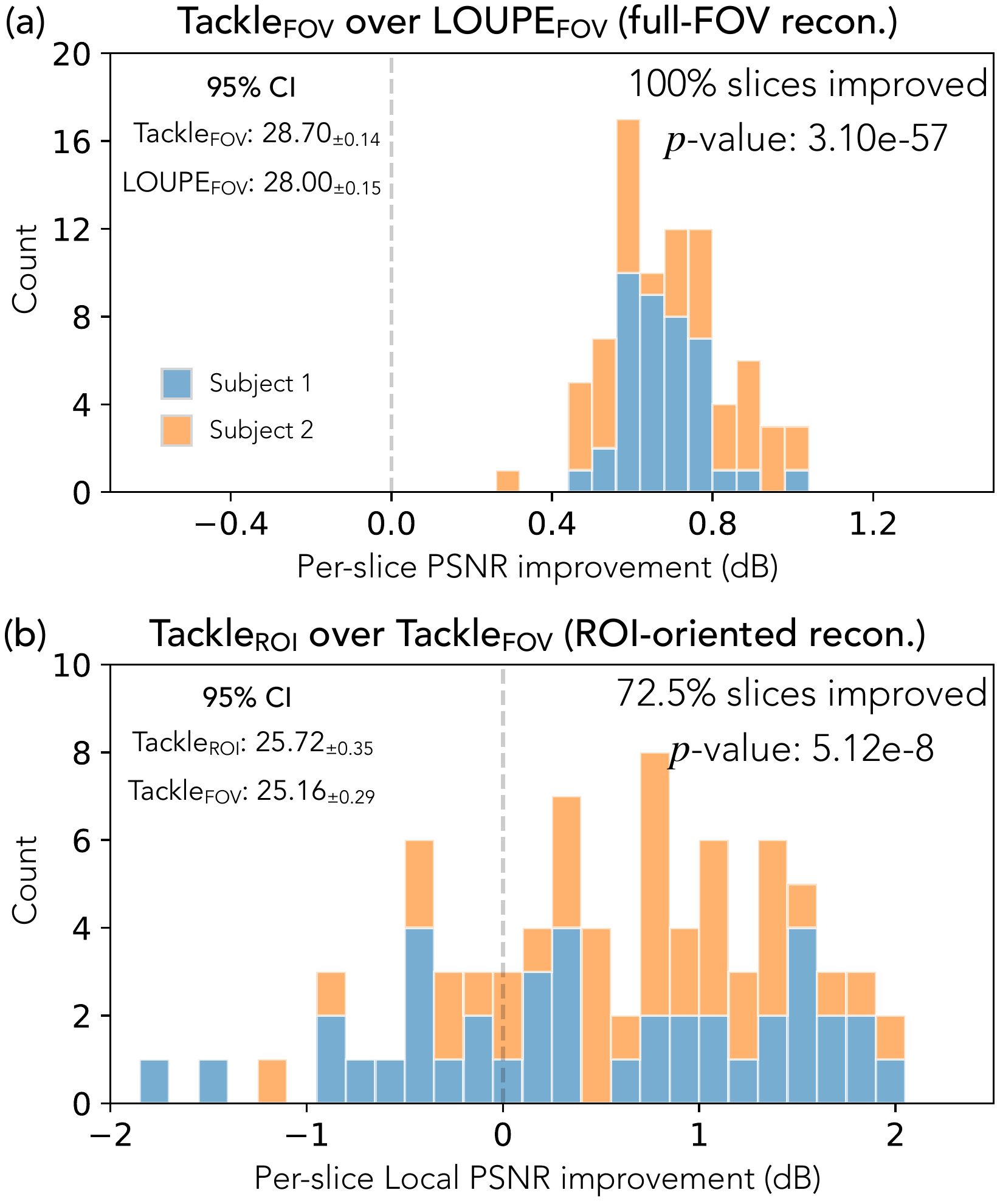}
\caption{
Slice-wise difference histograms. (a): \tacklefov~over \loupefov on the full-FOV reconstruction task and (b): \tackleroi~over \loupefov~on the ROI-oriented reconstruction task. 
The 95\% confidence intervals are given in the top left corner of each plot.
In both cases, the vast majority of slices are improved and the $p$-values given by the paired samples t-test are highly significant.
}
\lblfig{martinos_histograms}
\vspace{-10pt}
\end{figure}

\noindent \textbf{Generalization gaps} \quad
There are multiple generalization gaps between the training (fastMRI single-coil data) and test data:
\begin{itemize}
    \item \textit{Different hardware:} The acquired data are collected directly with a single-channel extremity coil, while the training data are simulated from $k$-space data collected by multi-channel receiver coils~\cite{knoll2020fastmri}.
    \item \textit{Different sequence and resolution:} The acquired data are given by a gradient-echo sequence with 1 mm isotropic resolution, while the training data are given by a spin-echo sequence with 0.5mm in-plane resolution~\cite{knoll2020fastmri}. 
    \item \textit{Different distribution of the ROI anatomy:} The acquired data are collected from two subjects whose menisci are healthy and have no tear, while the ROIs in the training data contain meniscus tears.
\end{itemize}
Despite these generalization gaps, \tackleroi~works robustly and leads to both numerical and visual improvement.

\noindent \textbf{Baselines} \quad In this section, we compare \tackleroi~with the following baselines under 4$\times$ acceleration.

\begin{itemize}
    \item \textit{Poisson-disc + Total Variation$_\text{FOV}$ (PD+TV$_\text{FOV}$):} The subsampling pattern is the same as the Poisson-disc sampling pattern generated by \texttt{sigpy.mri.poisson} for PD+UN$_\text{FOV}$ in \refsec{exp_roi}. The reconstruction is obtained by solving a total variation (TV) regularized optimization problem with the Sparse MRI toolbox\footnote{\href{https://people.eecs.berkeley.edu/~mlustig/Software.html}{https://people.eecs.berkeley.edu/~mlustig/Software.html}}.
    \item \textit{\loupefov:} the same \loupefov~baseline as in \refsec{exp_roi}.
    \item \textit{\tacklefov:} a \tackle~model trained for full-FOV reconstruction.
    \item \textit{\louperoi:} the same architecture as \loupefov~but trained for ROI reconstruction following the same training procedure as \tackleroi.
\end{itemize}

\begin{table}[t]
\scriptsize
\newcolumntype{C}{>{\centering\arraybackslash}p{34pt}}
\newcolumntype{D}{>{\centering\arraybackslash}p{52pt}}
\definecolor{LightGreen}{rgb}{0.78,96,0.71}
\centering
\caption{Comparison of average reconstruction accuracy on the experimentally collected dataset under 4$\times$ acceleration (top: full-FOV recon.; bottom: ROI-oriented recon.)}
\begin{tabularx}{248pt}{DCCCCC}
\toprule 
Full-FOV recon. & PD+TV$_\text{FOV}$ & \loupefov & \cellcolor{LightGreen} \tacklefov & \tackleroi \\ \midrule
PSNR (dB) & 27.94 & 28.00 & \textbf{28.70} & 28.18 \\ 
\bottomrule \\
\toprule
ROI recon. & PD+TV$_\text{FOV}$ & \loupefov & \tacklefov & \cellcolor{LightGreen} \tackleroi \\ \midrule
Local PSNR (dB) & 24.45 & 24.67 & 25.16 & \textbf{25.72} \\
\bottomrule
\multicolumn{5}{l}{\tikz \fill [LightGreen] (0,0.1) rectangle (0.5,0.25); indicates the variant of \tackle~with
matching training and evaluation metrics}
\end{tabularx}
\lbltab{exp_martinos}
\vspace{-10pt}
\end{table}

\begin{figure*}[t]
\centering
\includegraphics[width=0.95\textwidth]{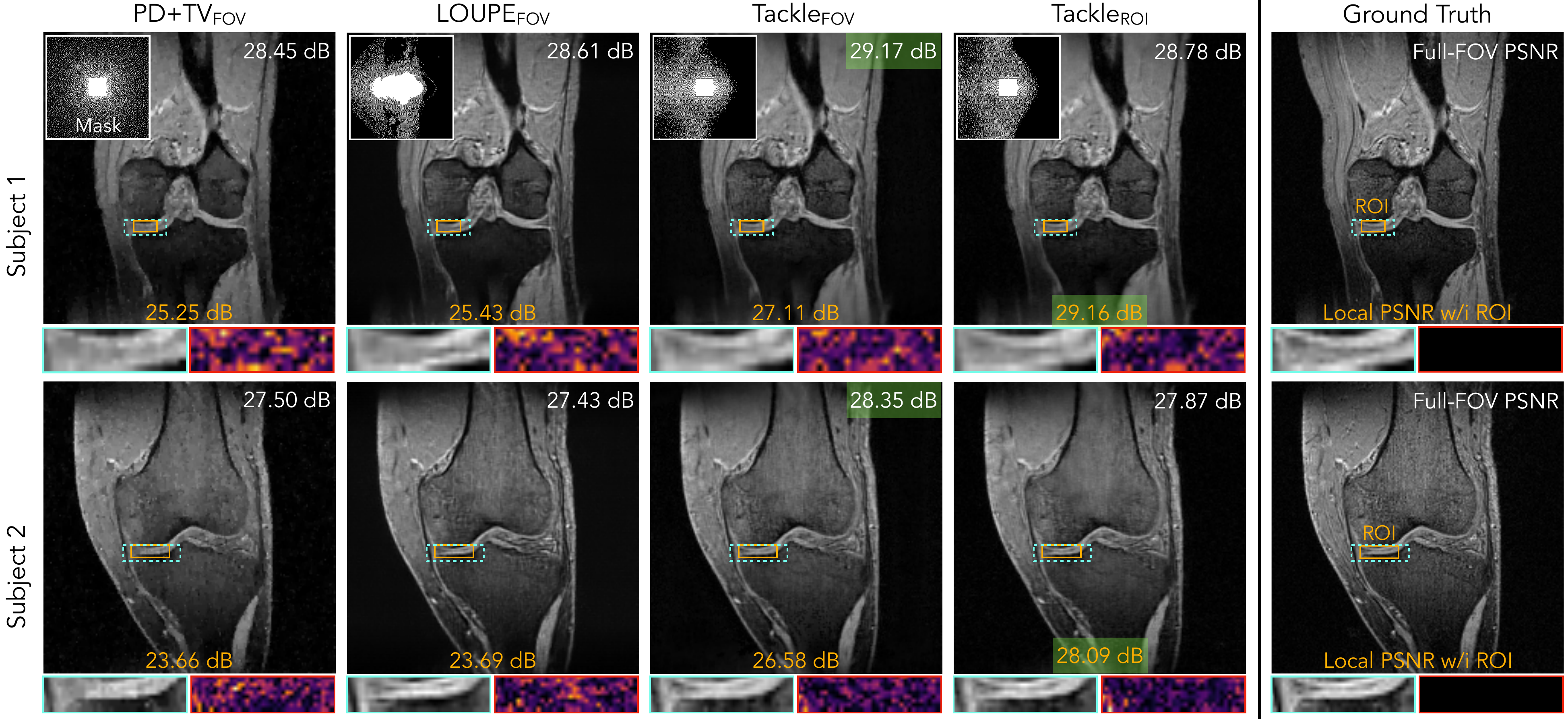}
\caption{
Reconstruction comparison of two samples in the experimentally collected dataset (top: from subject 1; bottom: from subject 2) by different methods under 4$\times$ acceleration. 
The sampling mask, a zoom-in on the ROI, and the error map are presented for each method. 
By sampling more frequencies along the vertical direction in $k$-space, \tackleroi~has a higher vertical resolution in the image space and thus outperforms other baselines optimized for full-FOV reconstruction on the ROIs with directional anatomical structure.}
\lblfig{martinos_examples}
\vspace{-10pt}
\end{figure*}

\noindent \textbf{Results} \quad
We present a quantitative comparison in \reftab{exp_martinos}.
For both the full-FOV and ROI-oriented reconstruction tasks, \tackle~outperforms the baselines under the corresponding metric.
For each task, we highlight the variant of \tackle~trained for the evaluation metric in green.
Our results show that the highlighted variant outperforms the other variant of \tackle, indicating a tradeoff between full-FOV and ROI reconstruction accuracy. 

We further conduct a slice-wise PSNR analysis in \reffig{martinos_histograms}. 
For both histograms, the horizontal axis is the improvement on the respective metric and the vertical axis is the count.
We also quantify the significance of the improvements using the paired samples t-test.
For the full-FOV reconstruction, \tacklefov~outperforms \loupefov~on \textit{all} 80 slices, giving a highly significant $p$-value of 3.10e-57.
We then compare \tackleroi~with the better full-FOV reconstruction method, \tacklefov, on the ROI-oriented reconstruction task.
Despite having the same architecture, \tackleroi~still outperforms \tacklefov~on 72.5\% of slices, leading to a $p$-value of 5.12e-8, which is also statistically significant. 
This result indicates that the ROI-oriented model \tackleroi~indeed provides more accurate ROI reconstructions on this out-of-distribution dataset.
We further provide some visual examples in \reffig{martinos_examples}. 
Below each reconstruction is a zoom-in on the region around the ROI and the error map of the region with respect to the ground truth.
\tackle~not only achieves higher PSNR values in both cases but also visually recovers the ROIs with fewer artifacts.

\noindent \textbf{Implementation} \quad
Besides the above results based on retrospective subsampling for quantitative comparison, we have also tested the learned sequence on a Siemens 3T MRI Skyra scanner.
Specifically, we implement a re-ordering loop that iterates through all the trajectories based on our learned subsampling mask $\mbm$.
The implemented sequence prospectively subsamples in $k$-space and shortens the scan time from 335 seconds to 84 seconds.
In \reffig{prospective}, we compare the reconstruction given by the prospectively subsampling sequence we implement with the reconstruction given by the retrospectively subsampled measurements from the fully sampling sequence. 
We note that the images labelled as ``\tackleroi~(retrospective)'' and ``\tackleroi~(prospective)'' are taken by two consecutive but separate scans, so there might be some subtle motion between them.
Nevertheless, the two images have no significant visual difference, indicating that the improvement we show on retrospective simulations translates into actual improvement in practice.
The prospective reconstruction successfully recovers important anatomical features around the meniscus region while only takes a quarter of the scan time compared to the full-sampled image.

\begin{figure}[t]
\centering
\includegraphics[width=0.48\textwidth]{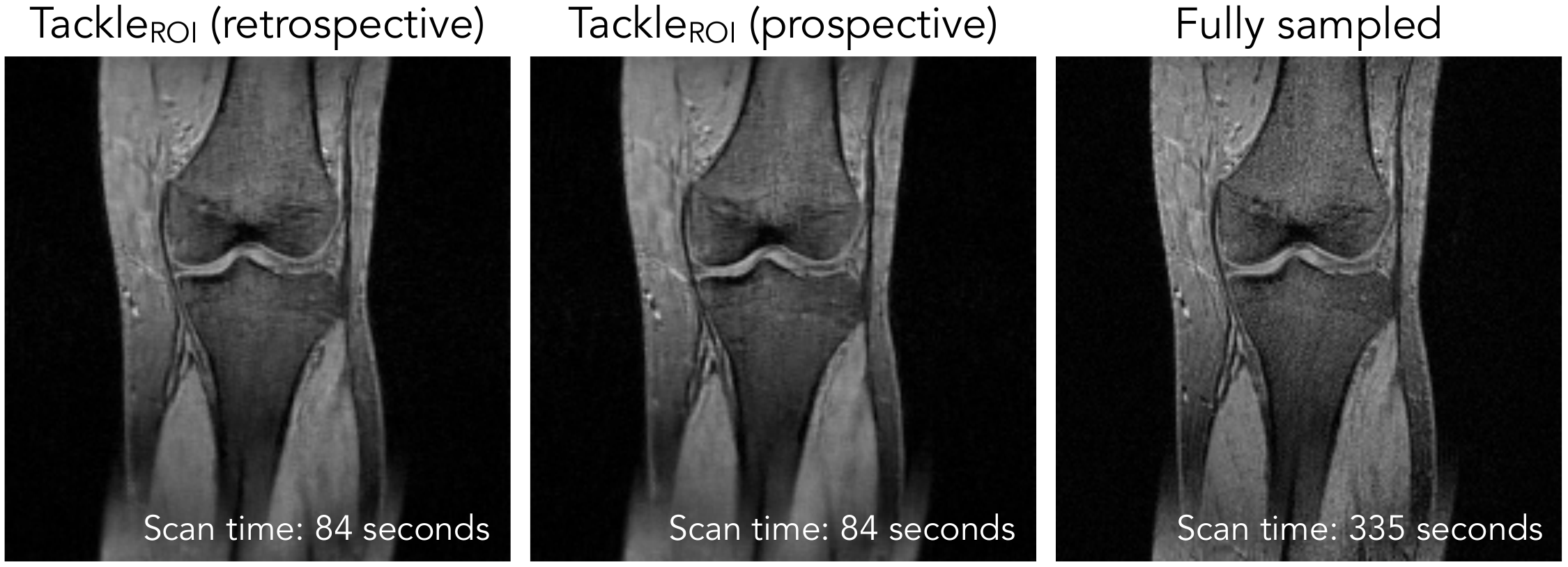}
\caption{
Reconstruction comparison between the implemented prospective subsampling sequence and the retrospective subsampling sequence. 
Our learned sequence can be implemented on an MRI scanner and generates images of quality indistinguishable from those recovered from retrospectively sampled data. 
Compared to the ground truth image, our prospectively subsampled reconstruction recovers important features around the meniscus region, which is the ROI it was trained to enhance.
}
\lblfig{prospective}
\vspace{-10pt}
\end{figure}

\vspace{-5pt}

\section{Ablation Studies}
\lblsec{ablation}

\subsection{Effectiveness of co-design}
\lblsec{ablation_codesign}

We evaluate the effectiveness of two aspects of co-design used in the proposed framework: learnable subsampling and task-specific training. 
In \reftab{ablation_codesign}, we compare four variants of the proposed method that have neither, one, or both aspects of co-design. 
The meanings of having or not having each aspect are summarized as follows:
\begin{itemize}
    \vspace{-2pt}
    \item Learnable subsampling (column 2)
        \begin{itemize}[label={},leftmargin=*]
        \vspace{-2pt}
            \item \xmark \, \textit{(Poisson-disc)}: use a Poisson-disc subsampling pattern that is randomly generated and then fixed
            \item \cmark: learn the subsampling pattern from data
        \end{itemize}
    \vspace{-2pt}
    \item Task-specific training (column 3)
        \begin{itemize}[label={},leftmargin=*]
        \vspace{-2pt}
            \item \xmark: separately optimize retriever and predictor
            \item \cmark: jointly optimize retriever and predictor
        \end{itemize}
\end{itemize}
To eliminate the effect of different network architectures, all four variants have exactly the same architectures.
Overall, we find both aspects of co-design beneficial.
For the task of reconstructing meniscus tear ROIs, learning the subsampling pattern is particularly helpful.
Task-specific training, on the other hand, is more important for the knee segmentation task.
Highlighted in cyan, the last row is the full-fledged version of \tackle, which achieves the best performance for all considered scenarios with both aspects of co-design.

\begin{table*}[t]
\scriptsize
\newcolumntype{C}{>{\centering\arraybackslash}p{40pt}}
\newcolumntype{D}{>{\centering\arraybackslash}p{25pt}}
\definecolor{LightCyan}{rgb}{0.88,1,1}
\centering
\caption{Ablation studies on two aspects of co-design for all the considered tasks under 16$\times$ acceleration}
\begin{tabularx}{494pt}{DccCCCCCC}
\toprule
\multirow{3}{*}{Method} & \multicolumn{2}{c}{\multirow{2}{*}{Ablated component}} & \multicolumn{2}{c}{ROI-oriented reconstruction} & \multicolumn{2}{c}{Tissue segmentation} & \multicolumn{2}{c}{Pathology classification} \\
 &  &  & \multicolumn{2}{c}{(Local PSNR in dB)} & \multicolumn{2}{c}{(Dice score)} & (Cls. acc.) & ($F_1$ score) \\
\cmidrule(lr){2-3} \cmidrule(lr){4-5} \cmidrule(lr){6-7} \cmidrule(lr){8-9}
 & Learned subsampling & Task-specific training & Single-coil & Multi-coil & Brain$^\P$ & Knee & \multicolumn{2}{c}{Gliomas tumor} \\ \midrule 
PD+VN$_\flat$ & \xmark \, (Poisson-disc) & \xmark & 29.91 & 36.48 & 0.9257 & 0.8018 & 0.9024 & 0.8871  \\
PD+VN$_\sharp$ & \xmark \, (Poisson-disc) & \cmark & 30.15 & 36.51 & 0.9256 & 0.8474 & 0.9072 & 0.8966  \\
\tackle$_\flat$ & \cmark & \xmark & 30.14 & 37.53 & 0.9350 & 0.8232 & 0.9062 & 0.8929  \\
\rowcolor{LightCyan}
\tackle$_\sharp$ & \cmark & \cmark & \textbf{31.54} & \textbf{37.89} & \textbf{0.9395} & \textbf{0.8532} & \textbf{0.9159} & \textbf{0.9039}  \\
\bottomrule
\multicolumn{9}{l}{$\flat$ \scriptsize{indicates full-FOV reconstruction oriented versions of PD+VN and \tackle \hspace{7cm} $\P$ see Supplement I}} \\
\multicolumn{9}{l}{$\sharp$ \scriptsize{indicates task-specific versions of PD+VN and \tackle}}
\end{tabularx}
\lbltab{ablation_codesign}
\vspace{-10pt}
\end{table*}

\vspace{-5pt}

\subsection{Effectiveness of the proposed architecture and training procedure}
\lblsec{ablation_architecture}

The proposed architecture of $\mathcal{T}_\theta$ from measurements $\ybm$ to prediction $\zbmhat$ consists of an E2E-VarNet retriever and a U-Net predictor.
A natural question is how this architecture compares with a single model-free neural network with a comparable number of parameters that directly maps subsampled measurements to the final prediction.
We consider the following comparisons in \reftab{ablation_archi}:
\begin{itemize}
    \vspace{-2pt}
    \item Single larger predictor (row 1)
        \begin{itemize}[leftmargin=*]
        \vspace{-2pt}
            \item \textit{Tissue seg.}: U-Net with 128 channels after the first convolution layer and the same number of pooling layers (42.2M parameters)
            \item \textit{Patho. class.}: ResNet101 (42.5M parameters)
        \end{itemize}
    \vspace{-2pt}
    \item VN+predictor (rows 2\&3)
        \begin{itemize}[leftmargin=*]
        \vspace{-2pt}
            \item \textit{Tissue seg.}: E2E-VarNet + standard U-Net (29.9M + 10.6M = 40.5M parameters)
            \item \textit{Patho. class.}: E2E-VarNet + ResNet18 (29.9M + 11.2M = 41.1M parameters)
        \end{itemize}
\end{itemize}
Comparing the first two rows, we find that the proposed ``VN+predictor'' architecture significantly outperforms the ``single larger predictor'' baseline on all settings.
This is likely due to the model-based nature of the ``VN+predictor'' architecture, which more effectively extracts useful information from subsampled measurements for downstream tasks.
Finally, we include the pre-training step discussed in \refsec{training_procedure}.
Highlighted in cyan, the full-fledged version of \tackle~in the last row significantly outperforms the ablated baselines on both non-reconstruction tasks, indicating the importance of both the proposed architecture and training procedure.

\begin{table}[t]
\scriptsize
\newcolumntype{B}{>{\centering\arraybackslash}p{28pt}}
\newcolumntype{C}{>{\centering\arraybackslash}p{32pt}}
\newcolumntype{D}{>{\centering\arraybackslash}p{24pt}}
\newcolumntype{E}{>{\centering\arraybackslash}p{44pt}}
\definecolor{LightCyan}{rgb}{0.88,1,1}
\centering
\caption{Ablation studies on model architecture and pre-training for non-reconstruction tasks under 16$\times$ acceleration}
\begin{tabularx}{256pt}{EBDDCC}
\toprule
\multicolumn{2}{c}{\multirow{2}{*}{Ablated component}} & \multicolumn{2}{c}{Tissue segmentation} & \multicolumn{2}{c}{Pathology classification} \\
 &  & \multicolumn{2}{c}{(Dice score)} & ($F_1$ score) & (Cls. acc.) \\
\cmidrule(lr){1-2} \cmidrule(lr){3-4} \cmidrule(lr){5-6}
Arch. of $\mathcal{T}_\theta$ & Pre-train & Brain$^\P$ & Knee & \multicolumn{2}{c}{Gliomas tumor} \\ 
\midrule 
Predictor only$^\ddag$ & \xmark & 0.9005 & 0.7539 & 0.8966 & 0.8788  \\
VN+predictor$^\ddag$ & \xmark & 0.9371 & 0.8163 & 0.9102 & 0.8969  \\
\rowcolor{LightCyan}
VN+predictor$^\S$ & \cmark & \textbf{0.9395} & \textbf{0.8532} & \textbf{0.9159} & \textbf{0.9039}  \\
\bottomrule 
\multicolumn{6}{l}{$\ddag$ \scriptsize{U-Net(128) / ResNet(101) for tissue seg. / patho. class.} \hspace{0.8cm} $\P$ see Supplement I} \\
\multicolumn{6}{l}{$\S$ \scriptsize{E2E-VarNet + U-Net(64) / ResNet(18) for tissue seg. / patho. class.}}
\end{tabularx}
\vspace{-10pt}
\lbltab{ablation_archi}
\end{table}

\vspace{-5pt}

\begin{table}[h]
\scriptsize
\newcolumntype{C}{>{\centering\arraybackslash}p{16pt}}
\newcolumntype{D}{>{\centering\arraybackslash}p{70pt}}
\definecolor{LightCyan}{rgb}{0.88,1,1}
\centering
\caption{Comparison of average test PSNR (dB) between reconstruction models trained with task-specific masks and \tacklerecon~on the fastMRI knee dataset}
\begin{tabularx}{250pt}{DCCCCCC}
\toprule
\multirow{3}{*}{Method} & \multicolumn{2}{c}{Brain seg.} & \multicolumn{2}{c}{Knee seg.} & \multicolumn{2}{c}{Tumor class.} \\
\cmidrule(lr){2-3} \cmidrule(lr){4-5} \cmidrule(lr){6-7}
 & 16$\times$ & 64$\times$ & 16$\times$ & 64$\times$ & 16$\times$ & 64$\times$ \\ \midrule 
Task-specific mask+VN & \textbf{38.47} & 33.04 & 32.53 & 30.10 & 44.20 & 37.07  \\
\tacklerecon & 38.44 & \textbf{33.13} & \textbf{32.63} & \textbf{30.24} & \textbf{44.48} & \textbf{37.26}  \\
\bottomrule
\end{tabularx}
\lbltab{task_mask_for_recon}
\end{table}

\subsection{Using task-specific sequences for reconstruction}
Our optimized task-specific pipeline learns to adjust the image representation from a conventional form to one that is more readily interpretable by the predictor network. 
This often adds additional textures to the images, making them look different from traditional reconstructions. 
However, this does not imply there is a significant loss in information that could be used for image reconstruction. 
Despite being optimized for task-specific objectives, our learned task-specific subsampling patterns can be used retrospectively for generating high-fidelity reconstructions. 
To show this, we conduct an experiment where we take the learned subsampling patterns of \tackleseg~and \tackleclass~and train an additional reconstruction network for each subsampling pattern.
The subsampling pattern is fixed during the training.
This experiment mimics the scenario if one wants a traditional reconstruction out of the collected $k$-space samples from our task-specific sequences.
In \reftab{task_mask_for_recon}, we provide a comparison with \tacklerecon, which jointly optimizes the subsampling pattern and reconstructor, on the fastMRI knee dataset.
One can see that the reconstruction models trained with task-specific masks (row 1) come close to \tacklerecon~(row 2) in terms of reconstruction performance.
These results indicate that our task-specific models do not incur a significant loss of image information but achieve a better trade-off for the downstream task accuracy.
It is thus possible to recover better images retrospectively using the $k$-space measurements collected by the task-specific sequences.

\section{Limitations}

Building on the promising results we have achieved, we acknowledge opportunities for further improvement of our current study.

\paragraph{Data usage} Similar to other works on task-specific CS-MRI co-design, our approach requires matched $k$-space, image, and annotation labels, which are of limited quantity in the research community.
Due to this limitation, two of our experiments (brain segmentation and tumor classification tasks) are conducted with $k$-space data simulated from magnitude images.

\paragraph{Sequence implementation} Although we have implemented a prospectively subsampling sequence with a learned sampling pattern by \tackleroi~on a Siemens MRI scanner, it was done using only one type of 3D gradient echo sequence. 
Other physical constraints affect the deployment of our method for general MRI sequences. 
For example, in spin-echo sequences, the order of sampling should be considered to mitigate spin-relaxation effects.

\paragraph{Controlled study} The evaluation in the current study is based on conventional quantitative metrics and qualitative visual comparisons. 
The number of volunteers for testing our learned sequences on a Siemens MRI scanner is relatively small.
To further assess prospective subsampling, future evaluations should involve controlled studies of image quality with radiologists.

\vspace{-5pt}

\section{Conclusion}
In this work, we generalized the objective of CS-MRI co-design to a variety of tasks beyond full-FOV reconstruction.
We introduced \tackle~as a unified approach for robustly learning task-specific strategies.
Through comprehensive experiments, we showed that \tackle~outperforms existing DL techniques that separately learn subsampling pattern, reconstruction, and prediction. Additionally, \tackle~outperforms na\"ive approaches to co-design that directly learn mappings from measurements to predictions.
We found that the optimized strategies sometimes circumvent the typical reconstruction in terms of pixel-wise accuracy, but effectively extract key visual information useful for task prediction.
Through ablation studies, we justified multiple design choices with regard to architecture and training procedure, and showed their importance in effectively learning CS-MRI strategies for tasks that go beyond full-FOV reconstruction.
We further implemented a learned subsampling sequence and tested it on a Siemens 3T MRI Skyra scanner, which led to a four-fold scan time reduction without sacrificing visual quality. 
Our study demonstrates the exciting promise of employing end-to-end co-design techniques, suggesting a future where clinical CS-MRI requirements are addressed with enhanced efficiency while maintaining accuracy.

\section*{Acknowledgment}

The authors would like to thank Xinyi Wu for her assistance as a volunteer in testing our learned MRI sequence and collecting data.

\bibliographystyle{IEEEtran}
\bibliography{./references}

\clearpage

\appendices
\renewcommand{\appendixname}{Supplement}

\resetcounter

\section{Brain tissue segmentation}
\lblapp{exp_seg}

In this section, we provide additional experimental results on a brain tissue segmentation problem in complement with the knee segmentation in the main paper.

\noindent \textbf{Dataset and setup} \quad
This study involves segmenting four brain tissues: the cortex, the white matter, the subcortical gray matter, and the cerebrospinal fluid (CSF). Following \cite{hoopes2021hypermorph}, we use the 109-th coronal slice of each full $k$-space sampled volume in the OASIS dataset \cite{marcus2007open} and the segmentation maps generated with SAMSEG in FreeSurfer \cite{fischl2012freesurfer}. 
SAMSEG, which stands for Sequence Adaptive Multimodal SEGmentation, is an established method for brain tissue segmentation and is considered a standard method for this task~\cite{puonti2016fast}.
We use the segmentation maps generated by SAMSEG as the supervised labels for training.
We use the same measurement simulation procedure as in the tumor classification experiments.
We simulate the single-coil $k$-space data for each image by taking the Fourier transform of the image and adding complex additive white Gaussian noise (AWGN), according to the forward model in Equation (1). 
The standard deviation of the noise for each image is 0.05\% of the magnitude of the DC component. 
We train each method to minimize the Dice loss until convergence and select the model with the highest Dice score on the validation set.

\noindent \textbf{Baselines} \quad
We compare \tackleseg~with the same baselines as the ones in the knee tissue segmentation experiments: \louperecon, PD+UN$_\text{recon.}$, \tacklerecon, and SemuNet.

\noindent \textbf{Results} \quad
We first provide a numerical comparison in \reftab{exp_oasis} and a boxplot comparison in \reffig{brain_box}.
Within the rectangle between each pair of methods in \reffig{brain_box}, the top number is the percentage of improved samples and the bottom number is the $p$-value given by the paired samples t-test.
With an improved architecture, \tacklerecon~significantly outperforms the other reconstruction-oriented baselines.
Nevertheless, \tackleseg~still outperforms \tacklerecon~under both accelerations with significant $p$-values, highlighting the benefit of task-specific training.
Compared to SemuNet \cite{wang2021one}, \tackleseg~learns better segmentation strategies for both acceleration ratios and is more robust to high acceleration. 
We further provide some visual examples in \reffig{brain_examples}, visualizing the input and output of the predictor across different methods.
The zoom-in regions highlight a location where the segmentation prediction of \tackleseg~outperforms other baselines. 
Specifically, \tackleseg~more accurately predicts the outline of the white matter (in yellow) than other methods.
Such an improvement leads to more precise estimation of the thickness of the cortex (in orange), an important task for studying human cognition and neurodegeneration \cite{apostolova2007brain}.

\section{Implementation details}
\lblapp{imple}

In this section, we describe the implementation details of \tackle~and the baseline methods.

\subsection{Further information on datasets and their preparation}

For each dataset in Section IV of the main text, we randomly split the data into training, validation, and test sets \textit{on the patient level}, which means that each validation or test slice comes from a patient whose images are not used for training. 

\subsubsection{ROI-oriented reconstruction}
For this task, we use all images with Meniscus Tear (MT) annotations in the fastMRI+ dataset \cite{zbontar2018fastMRI, zhao2022fastmri+}.
We follow the specific data splitting in \cite{pineda2020active}, which results in 4,158 slices for training, 210 slices for validation, and 201 slices for testing. 
We crop the center of the $k$-space of each image and adjust the size and position of each bounding box accordingly.

\subsubsection{Brain tissue segmentation}
For this task, we use the 109-th coronal slice of each volume in the OASIS dataset \cite{marcus2007open}.
The access to the dataset can be found here\footnote{\href{https://github.com/adalca/medical-datasets/blob/master/neurite-oasis.md}{https://github.com/adalca/medical-datasets/blob/master/neurite-oasis.md}}.
Specifically, we use the 4-label tissue-type segmentation maps, which include segments of the cortex, the white matter, the subcortical gray matter, and the cerebrospinal fluid (CSF).
We split the data into 248 slices for training ($\approx$60\%), 82 slices for validation ($\approx$20\%), and 84 slices for testing ($\approx$20\%). 

\subsubsection{Knee tissue segmentation}
For this task, we use all the sagittal slices in the SKM-TEA dataset  \cite{desai2021skmtea} that contains all four segmentation labels (the patellar cartilage, the femoral cartilage, the tibial cartilage, and the meniscus).
We split the data into 2,935 slices for training ($\approx$60\%), 1,040 slices for validation ($\approx$20\%), and 987 slices for testing ($\approx$20\%).

\subsubsection{Pathology classification}
For this task, we use all the images acquired by the FLAIR sequence in the BRATS dataset \cite{menze2014multimodal} to detect the existence of the Glioma tumor.
FLAIR stands for \textit{fluid attenuated inversion recovery}, a kind of inversion recovery sequence that is commonly used for detecting various brain lesions due to its ability of suppressing the CSF signal and enhancing lesion-to-background contrast \cite{kates1996fluid}.
Empirically, we find that it is more accurate to detect the existence of the Glioma on FLAIR images than on images with the other contrasts in the BRATS dataset.
We split the data into 30,495 slices for training ($\approx$60\%), 9,996 slices for validation ($\approx$20\%), and 10,353 slices for testing ($\approx$20\%).

\begin{table}[t]
\centering
\scriptsize
\caption{Comparison of average test Dice score on the brain segmentation task under different acceleration ratios ($R$)}
\begin{tabular}{cccccc}
\toprule 
$R$ & PD+UN$_\text{recon.}$ & \louperecon & SemuNet & \tacklerecon & \tackleseg \\ \midrule
16 & 0.8952 & 0.9244 & 0.9196 & 0.9350 & \textbf{0.9395} \\
64 & 0.8377 & 0.8733 & 0.3824 & 0.9181 & \textbf{0.9218} \\
\bottomrule
\end{tabular}
\lbltab{exp_oasis}
\end{table}

\begin{figure}[t]
\centering
\includegraphics[width=0.45\textwidth]{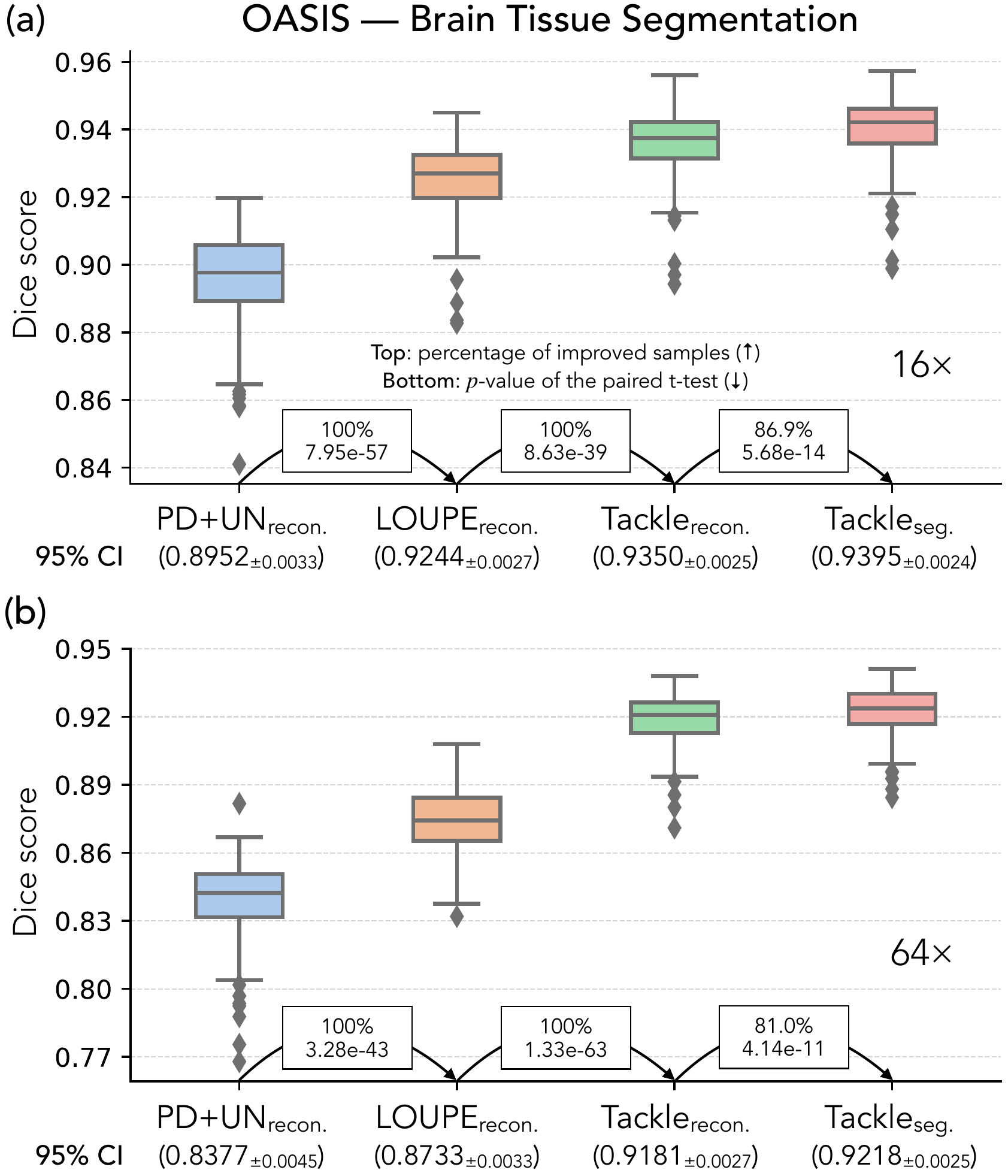}
\caption{
Box plots of the brain tissue segmentation results under 16$\times$ (a) and 64$\times$ (b) accelerations. 
Within the rectangle between each pair of methods, the top number is the percentage of samples that get improved and the bottom number is the $p$-value given by the paired samples t-test.
A higher percentage and lower $p$-value indicate a more significant improvement.
We also provide the 95\% confidence intervals for all methods below their names.
Similar to the knee segmentation results, the proposed method \tackle~outperforms other baselines in terms of all the statistical measures for both acceleration ratios.
}
\vspace{-10pt}
\lblfig{brain_box}
\end{figure}

\begin{figure*}[t]
\centering
\includegraphics[width=0.95\textwidth]{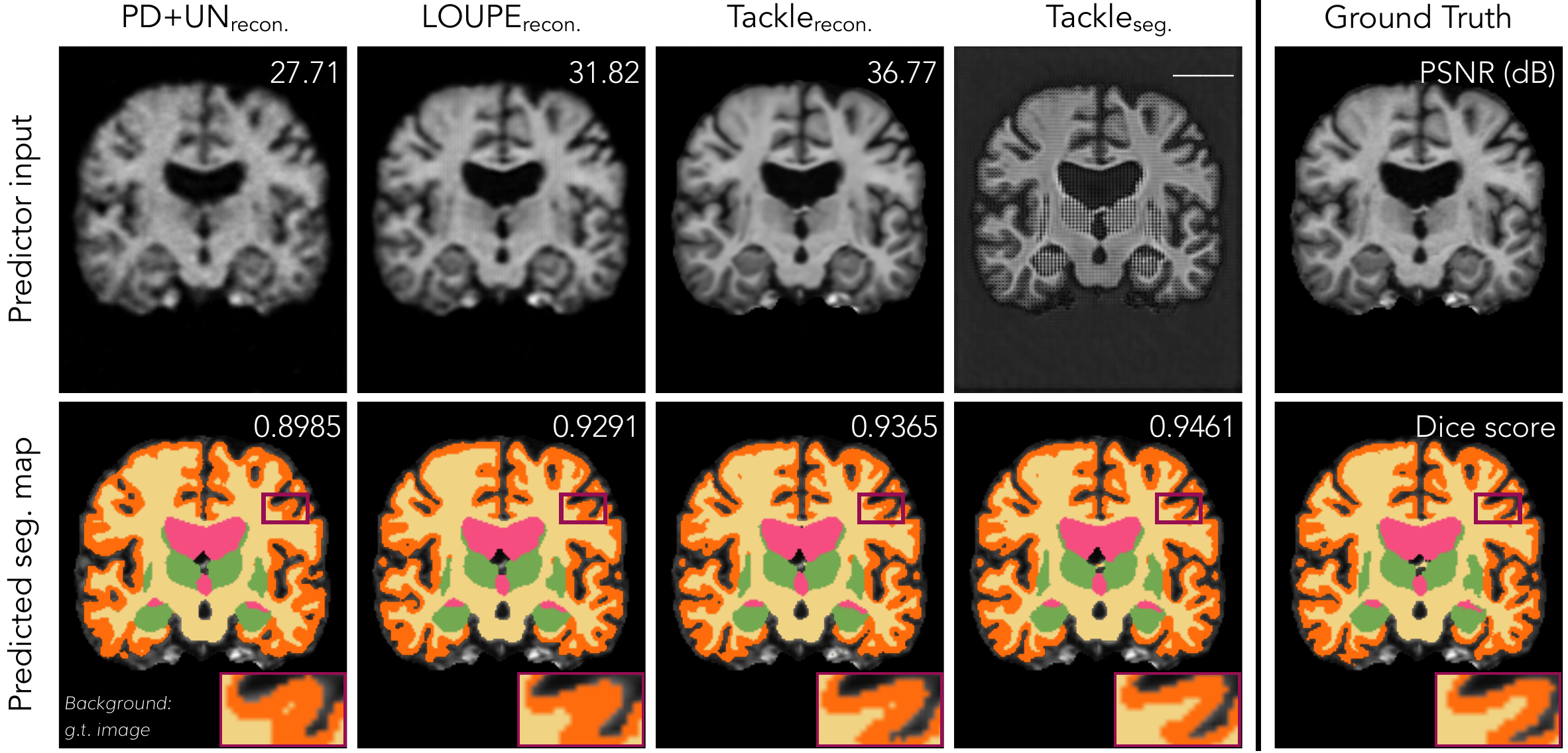}
\caption{
Comparison of segmentation results under 16$\times$ acceleration on one sample from the OASIS dataset. 
Similar to the knee segmentation results, \tackleseg~circumvents the typical ``reconstruction'' in terms of pixel-wise similarity with the ground truth image.
Instead, it learns an anatomically accurate feature map, which enables better segmentation prediction than other baselines both for this sample and on average over the test set (\reftab{exp_oasis}).
The zoom-in panels highlight a region where \tackleseg~more accurately predicts the outline of white matter (in yellow) than other methods.
This improvement leads to a more precise estimation of the thickness of the cortex (in orange), an important task for studying human cognition and neurodegeneration \cite{apostolova2007brain}.
}
\vspace{-10pt}
\lblfig{brain_examples}
\end{figure*}

\subsection{Training and implementation details of \tackle}

\subsubsection{Training details}
For all experiments, we train the model using the Adam \cite{kingma2014adam} optimizer with $\beta_1 = 0.9, \beta_2 = 0.999$ on a single NVIDIA A6000 GPU. 
We choose the best learning rate among \{1e-2, 1e-3, 1e-4\}, and trained all models until convergence (i.e. no improvement for 10 epochs on the validation set according to the task-specific evaluation metric). 
For instance, if a \tackleseg~model achieves a higher Dice score on the validation set than all previous epochs on epoch 42, the model will be saved as a checkpoint.
If it has no further improvement until epoch 52, then the training will be terminated and the saved checkpoint on epoch 42 will be used for reporting the final results.

The training of our proposed framework is conducted by retrospective subsampling on fully sampled measurements.
The first module is the sampler, which requires no input and directly learns a matrix that contains the probability of sampling each $k$-space frequency.
The output of the sampler is the subsampling mask $\mbm$, in which 1 represents the measurements to be sampled and 0 represents those not to be sampled.
Sampling amounts to taking the element-wise product between $\mbm$ and the fully sampled measurements $\kbm$, which gives us the subsampled measurements $\ybm := \mbm \odot \kbm$. 
The retriever will then take the two-channel complex measurements $\ybm$ as the input and output a single-channel real image $\xbmhat$.
In the multi-coil case, $\ybm$ contains signals from multiple coils with different sensitivity maps and $\xbmhat$ is reconstructed by taking the root sum of square across all coils.
For reconstruction tasks (full-FOV reconstruction and ROI-oriented reconstruction), $\zbmhat = \xbmhat$ will be the final output for loss calculation and back-propagation.
For downstream tasks beyond reconstruction, we feed $\xbmhat$ into an additional predictor which gives on a prediction $\zbmhat$.
In this case, $\zbmhat$ will be the final output for loss calculation and back-propagation.

\subsubsection{Retriever architecture}
Following the E2E-VarNet architecture \cite{sriram2020end}, our retriever operates in $k$-space and contains 12 refinement steps, each of which includes a U-Net \cite{ronneberger2015u} with independent weights from each other. 
The update rule of the $t$-th refinement step is
$$\kbm^{t+1}=\kbm^t-\eta^t \diag(\mbm) \left(\kbm^t-\ybm\right)+\mathsf{G}^t \left(\kbm^t\right)$$
where $\mbm$ is the subsampling mask, $\ybm$ is the measurement, $\kbm^t$ is the reconstructed $k$-space, $\eta^t$ is a data consistency parameter, and $\mathsf{G}^t$ is the refinement module defined as
$$\mathsf{G}^t \left(\kbm^t\right):=\Fbm \Ebm \left( \mathsf{UN}^t \left( \Rbm \Fbm^{-1}\kbm^t \right) \right).$$
Here, $\Ebm$ and $\Rbm$ are the expand and reduce operations across all coils (see \cite{sriram2020end} for more details), and $\mathsf{UN}^t$ is the U-Net model at $t$-th step.
Specifically, we use the standard U-Net \cite{ronneberger2015u} architecture with 2 input and output channels, 4 average down-pooling layers, and 4 up-pooling layers. 
The model starts with a 18-channel output for the input layer and doubles the number of channels with each downsampling layer.
Between every two pooling layers are two convolution modules, each of which consists of a $3\times 3$ convolution, an instance normalization \cite{ulyanov2016instance}, and a LeakyReLU activation with negative slope of 0.2.
The input to each U-Net is first normalized to zero mean and standard deviation of 1 before being fed into the network, and will be normalized back to the original mean and standard deviation after passing through the network.
After 12 refinement steps, the final output layer of the retriever is an inverse Fourier transform followed by a root-sum-squares reduction for each pixel over all coils.
The output of the retriever is a batch of single-channel images.
For reconstruction tasks, a loss function will be directly applied to the output.
For non-reconstruction tasks, there is an additional predictor module.

\subsubsection{Predictor architecture}
For tissue segmentation tasks, the predictor is a U-Net model that has the same architecture as the refinement network described above except for the following differences: 
There are 1 input channel and $c$ output channels (where $c$ is the number of segmentation classes). 
The model starts with a 64-channel output for the input layer.
The convolution modules use the Parametric ReLU activation.
There is no normalization after the output.
We used the U-Net implementation in the \texttt{MONAI} package \cite{cardoso2022monai}.
For the pathology classification task, the predictor is a ResNet18 model with 1 input channel and 2 output dimensions.
We normalize the input to zero mean and standard deviation of 1 before feeding it into the network.
We used the ResNet implementation in the \texttt{torchvision} package \cite{marcel2010torchvision}.

\subsection{Pre-select region and sensitivity map estimation}
Among all the datasets considered in this manuscript, fastMRI+ \cite{zbontar2018fastMRI, zhao2022fastmri+} and SKM-TEA \cite{desai2021skmtea} contain multi-coil $k$-space measurements.
Reconstruction from multi-coil $k$-space data requires estimation of the coil sensitivity maps, i.e. $\Sbm_i$ in Equation (2) of the main text, using the central low-frequency region of the $k$-space, called the Auto-Calibration Signal (ACS).
We set the ACS region as a square around the DC component that contains $\nicefrac{1}{8}$ of the subsampling budget. 
For example, if a dataset contains $k$-space measurements of size $256 \times 256$, for 8$\times$ acceleration, we will select the center $32 \times 32$ low frequencies as the ACS.
We also include the pre-determined ACS region for single-coil $k$-space experiments because we find that it stabilizes the training of some baselines.

Given the ACS, we estimate coil sensitivity maps using the Sensitivity Map Estimation (SME) module introduced in \cite{sriram2020end}.
In contrast to the ESPIRiT algorithm \cite{uecker2014espirit}, SME estimates the sensitivity maps with a CNN applied to each coil image independently.
The architecture of the CNN in SME is the same as the U-Net in each E2E-VarNet cascade, except with an 8-channel output instead of an 18-channel output for the input layer.

\subsection{Further details on the implementation of SemuNet \cite{wang2021one}}
For the brain and knee segmentation tasks, we compare the proposed method with SemuNet \cite{wang2021one}. 
Originally demonstrated for a brain segmentation task, it also aims to jointly optimize a subsampling mask, a reconstructor, and a task predictor for the downstream accuracy.
SemuNet uses a hybrid of $\ell_1$ loss for reconstruction and cross-entropy loss for segmentation.
Since the code of SemuNet is not released, we have tried to reproduce the results in the original paper to our best efforts.
Specifically, we follow their proposed loss function and architecture of the sampler, the residual U-Net reconstructor, and a U-Net predictor.
We follow the original paper to use an Adam optimizer \cite{kingma2014adam} and not pre-select low-frequency measurements.
However, since our tasks and datasets are different from those in \cite{wang2021one}, we empirically find that the learning rate and the parameter $\lambda$ that adjusts the trade off between the two losses are suboptimal for our settings. 
Therefore, we conduct a grid search on the learning rate in $\{0.001, 0.005, 0.01, 0.05, 0.1, 0.5, 1, 5\}$ and $\lambda \in \{0.0001, 0.001, 0.01\}$.
For both $16\times$ and $64\times$ accelerations, we choose the best combination of parameters based on the performance on the validation set, and report the Dice score on the test set.

\section{Subsampling setup and implementation}
\lblapp{setup}

\begin{figure}[t]
\centering
\includegraphics[width=0.45\textwidth]{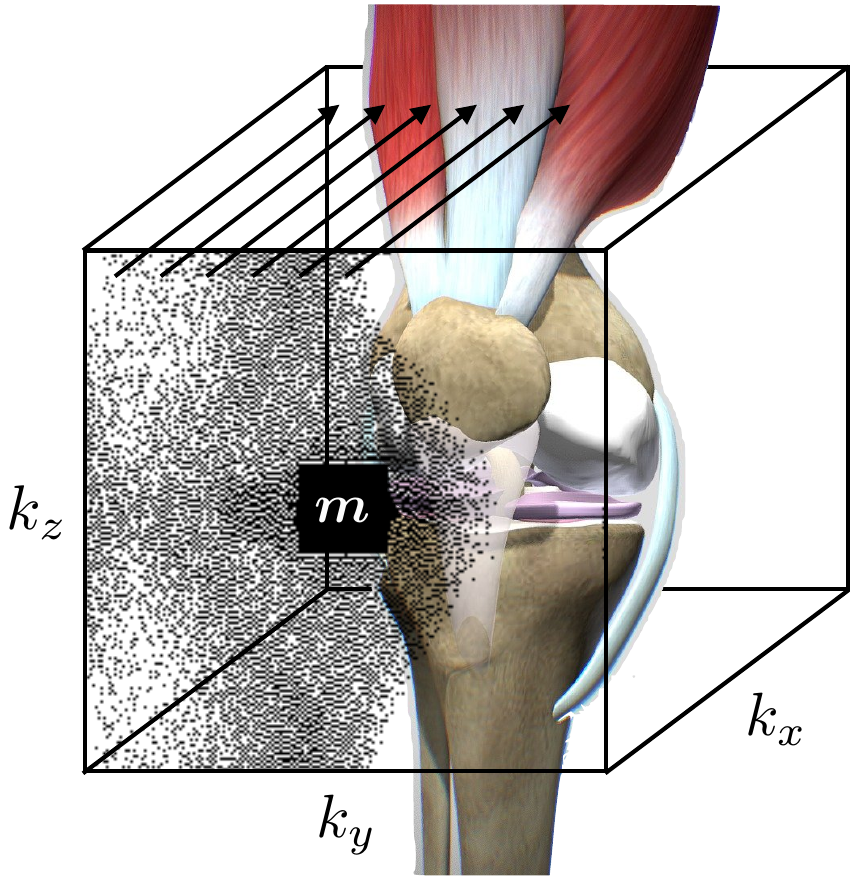}
\caption{
Conceptual illustration of the subsampling setup with a knee example. The back dots on the $k_y$-$k_z$ plane represents $k$-space trajectories along $k_x$, which are illustrated by the black arrows. We consider subsampling in the two phase-encoding dimensions ($k_y$ and $k_z$) of a 3D Cartesian sequence, where the subsampling pattern $\mbm$ is learned from data for some specific downstream task.}
\lblfig{implementation}
\vspace{-10pt}
\end{figure}

In this work, we optimize the subsampling mask $\mbm$ over all 2D subsampling patterns.
We implement 2D subsampling patterns in practice by subsampling in the two phase encoding dimensions of a 3D Cartesian sequence based on the 2D pattern, as illustrated in \reffig{implementation}. 
We denote the number of trajectories along $k_y$ and $k_z$ (the two phase encoding directions) as $n_{k_y}$ and $n_{k_z}$, respectively. 
For the fully sampling scenario, one needs to sequentially sample a total of $n_{k_y}n_{k_z}$ trajectories, which could take a long time to acquire in practice. 
Given a 2D subsampling mask $\mbm$, we subsample in the $k_y$-$k_z$ plane according to $\mbm$.
If $\mbm$ has an acceleration ratio of $R$, the subsampling sequence only takes $n_{k_y}n_{k_z}/R$ trajectories and the acquisition time will be reduced by a factor $R$ in practice. 
One can obtain the slice-wise 2D $k$-space measurements $\ybm$ by taking the 1D inverse Fourier Transform of the raw 3D $k$-space data along $k_x$.
We have implemented a $4\times$-accelerated version of the sequence that we used in Section V of the main text on the Siemens IDEA sequence programming platform, using the subsampling scheme we described above. 
Figure 10 in the main text demonstrates that the prospective subsampling version of our learned sequence achieves the same level of visual quality as the fully sampling version but only takes a quarter of the scan time.
This result highlights the real-world practicality of our approach.

\section{Additional validation on out-of-distribution data}

In the main text, we showed that \tackleroi~improves the reconstruction of ROIs that contain the meniscus tear (MT). 
In practice, it is likely that a healthy subject or someone with a different pathology lesion from the meniscus tear will get scanned.
So it is important that the learned sequence should also generalize to out-of-distribution subjects.
Here we take our trained \tackleroi~models with 4$\times$, 8$\times$, and 16$\times$ accelerations from our ROI reconstruction experiments, and directly test them on images that do not contain the meniscus tear, without additional fine-tuning.
The results are summarized in \reftab{out_of_distribution}.

\begin{table}[t]
\centering
\caption{Comparison of \tackleroi~on in- and out-of-distribution samples under different acceleration ratios ($R$)}
\begin{tabular}{cccc}
\toprule 
\multirow{2}{*}{Metric} & \multirow{2}{*}{$R$} & Samples w/ MT & Samples w/o MT   \\ 
 &  & (in-distribution) & (out-of-distribution)   \\ \midrule
 & 4 & 37.65 & 36.92  \\
PSNR (dB) & 8 & 33.28 & 32.88  \\
 & 16 & 32.06 & 31.74  \\
\bottomrule
\lbltab{out_of_distribution}
\vspace{-10pt}
\end{tabular}
\end{table}

Although it is not surprising that \tackleroi~performs better on the in-distribution data (samples w/ MT), we want to point out that the two numbers above correspond to two different test sets and are thus not directly comparable.
The main takeaway is that \tackleroi~can robustly recover samples without MT, even if it is trained on samples with MT. 
As discussed in the Section IV.A of the main text, \tackleroi~improves the ROI reconstruction by trading off $k$-space frequencies for the local anatomy to attain improved resolution.
We find that such a strategy can lead to satisfactory reconstruction quality even when the underlying subject does not contain the target pathology.

\section{Additional results}

\begin{figure}[t]
\centering
\includegraphics[width=0.45\textwidth]{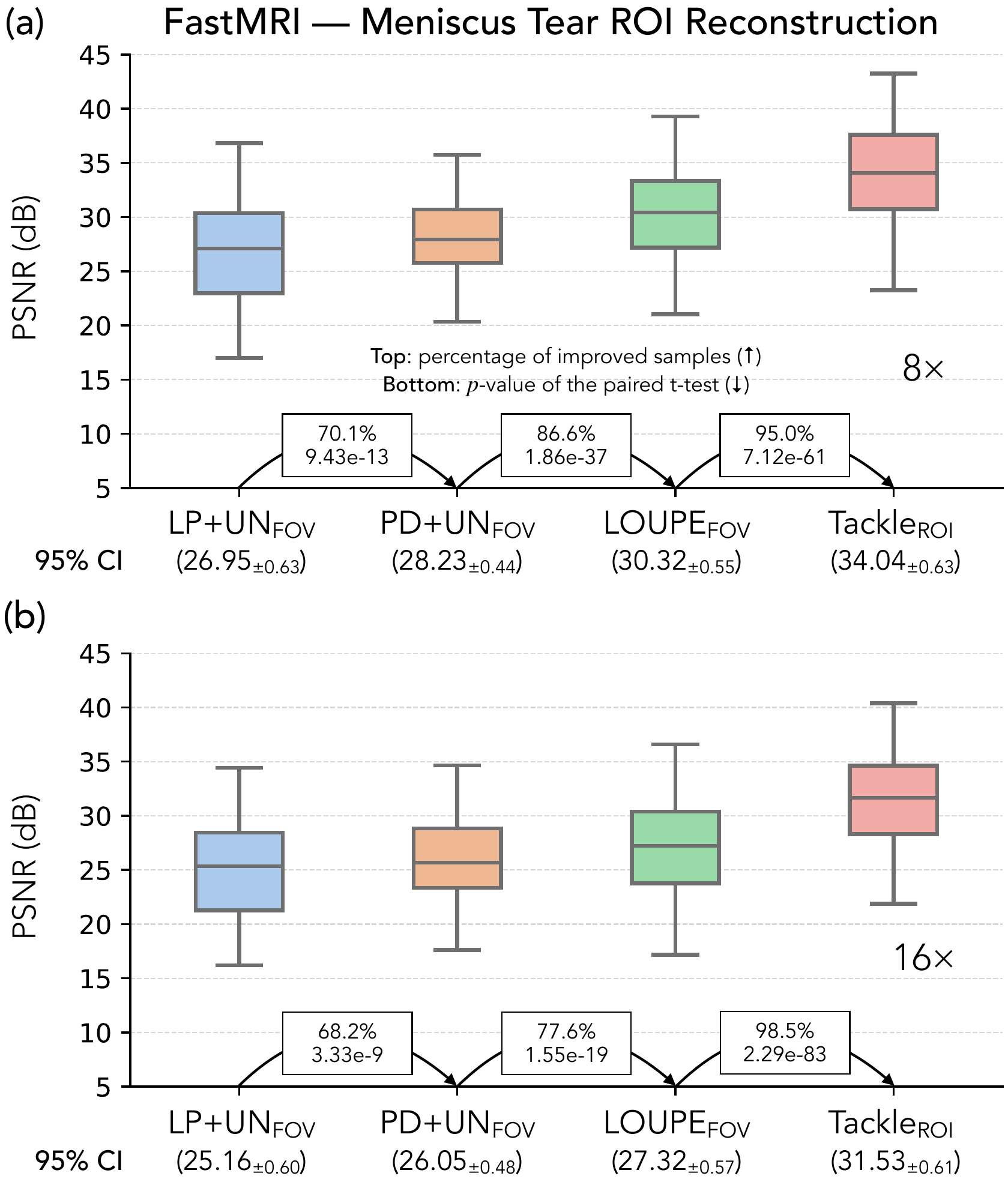}
\caption{
Box plots of the Meniscus Tear ROI reconstruction results under 8$\times$ (a) and 16$\times$ (b) accelerations. 
Within the rectangle between each pair of methods, the top number is the percentage of samples that get improved and the bottom number is the $p$-value given by the paired samples t-test.
A higher percentage and lower $p$-value indicate a more significant improvement.
The 95\% confidence intervals for all methods are given below their names.
}
\vspace{-10pt}
\lblfig{roi_box}
\end{figure}

We provide a boxplot comparison for the Meniscus Tear ROI reconstruction in \reffig{roi_box}.
We also provide some visual examples for the tumor classification task in \reffig{cls_examples}. 
For the three reconstruction-oriented baselines (left three columns), the inputs to the predictor network are typical reconstructions.
Optimized end-to-end for classification accuracy, the retriever of \tackleclass~learns a feature map that highlights the region where a tumor could exist.
Similar to the segmentation results, we find that the end-to-end model \tackleclass~circumvents the typical reconstruction but preserves image-level features that are helpful for downstream classification prediction.

\begin{figure*}[t]
\centering
\includegraphics[width=0.95\textwidth]{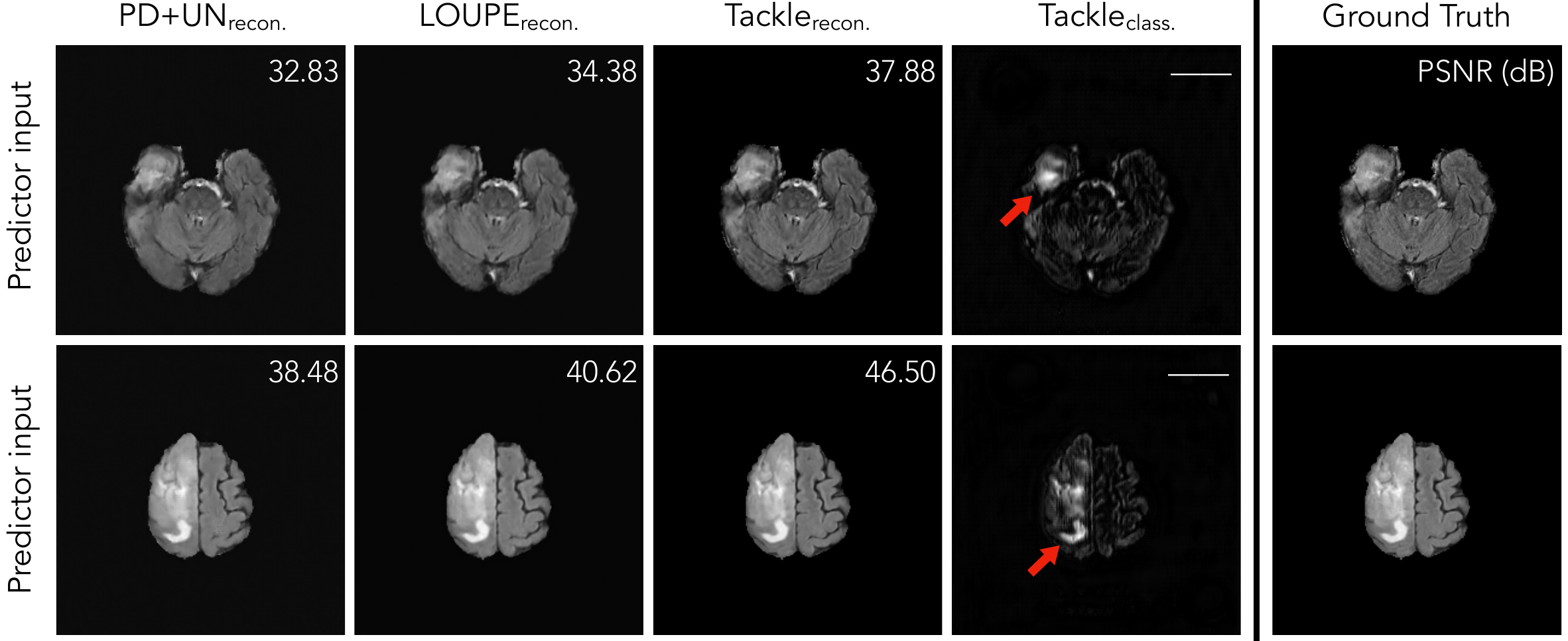}
\caption{
Visualization of the input of the predictor network for the brain tumor classification task under 16$\times$ acceleration on two samples of the BRATS dataset. 
Similar to the segmentation results, as a co-design method, \tackleclass~circumvents the typical ``reconstruction'' in terms of point-wise similarity with the ground truth image.
Instead, the retriever learns a feature map that highlights the region around the tumor for the downstream prediction.
}
\lblfig{cls_examples}
\vspace{-10pt}
\end{figure*}

\end{document}